\documentclass[onecolumn,11pt,showpacs,preprintnumbers,amsmath,amssymb]{revtex4}

\usepackage{graphicx}
\usepackage{dcolumn}
\usepackage{bm}
\usepackage{subfigure}
\usepackage{marvosym}
\usepackage{textcomp}
\usepackage{threeparttable}
\usepackage{booktabs}
\usepackage{setspace}
\usepackage{latexsym}
\renewcommand{\TPTtagStyle}%
{\normalsize\textit}

\begin{document}

\title{Holographic Photon Production with Magnetic Field in Anisotropic Plasmas}
\author{Shang-Yu Wu$^{1,2,3}$\footnote{loganwu@gmail.com}, Di-Lun Yang$^4$\footnote{dy29@phy.duke.edu}}
\affiliation{$^1$Institute of physics, National Chiao Tung University, Hsinchu 300, Taiwan.\\
$^2$National Center for Theoretical Science, Hsinchu, Taiwan.\\
$^3$Yau Shing Tung Center, National Chiao Tung University, Hsinchu, Taiwan\\
$^4$Department of Physics, Duke University, Durham, North Carolina 27708, USA}%
\date{\today}

\begin{abstract}
We investigate the thermal photon production from constant magnetic field in a strongly coupled and anisotropic plasma via the gauge/gravity duality. The dual geometry with pressure anisotropy is generated from the axion-dilaton gravity action introduced by Mateos and Trancancelli and the magnetic field is coupled to fundamental matters(quarks) through the D3/D7 embeddings. We find that the photon spectra with different quark mass are enhanced at large frequency when the photons are emitted parallel to the anisotropic direction with larger pressure or perpendicular to the magnetic field. However, in the opposite conditions for the emitted directions, the spectra approximately saturate isotropic results in the absence of magnetic field. On the other hand, a resonance emerges at moderate frequency for the photon spectrum with heavy quarks when the photons move perpendicular to the magnetic field. The resonance is more robust when the photons are polarized along the magnetic field. On the contrary, in the presence of pressure anisotropy, the resonance will be suppressed. There exist competing effects of magnetic field and pressure anisotropy on meson melting in the strongly coupled super Yang-Mills plasma, while we argue that the suppression led by anisotropy may not be applied to the quark gluon plasma.
\end{abstract}

\maketitle
\section{\label{sec:level1}Introduction}
In the relativistic heavy ion collisions, electromagnetic signatures may provide imperative information about the local properties of the quark gluon plasma(QGP) or even the early-time physics before thermalization due to their weak interaction with the medium. In comparison with p-p collisions, the enhanced yields of "direct photons" (i.e. not originating from the hadronic decays) observed in RHIC may indicate the prominent thermal-photon production from the QGP\cite{Adare:2008ab,Adare:2009qk}. In general, the photons with large energy (transverse momentum $p_T>1$ GeV) are expected to be generated in early times. Except for the yields of photons, the elliptic flow $v_2$ as a coefficient of the second Fourier harmonic of the particle spectrum is used to analyze the momentum-space anisotropy of produced photons. The explicit definition of $v_2$ can be found in \cite{Gale:2009gc}. Intuitively, the $v_2$ for direct photons is expected to be small since the flow should be gradually built up in the hydrodynamic phase. However, the recent results from both Au-Au and Pb-Pb collisions in RHIC and LHC have shown that the energetic photons carry large $v_2$ \cite{Adare:2011zr,Lohner:2012ct}, which contradict the theoretical expectation and may suggest an early-time mechanism contributing to the anomalous flow.   

On the theory side, many efforts have been devoted to the thermal-photon production, which leads to considerable contributions to direct photons, in the weakly coupled plasma \cite{Kapusta:1991qp,Baier:1991em,Aurenche:1998nw,Arnold:2001ms,Arnold:2002ja,Ghiglieri:2013gia}. 
Nevertheless, due to the strongly coupled feature of QGP at finite temperature, the non-perturbative approaches are also required. The AdS/CFT correspondence\cite{Maldacena:1997re,Witten:1998qj,Gubser:1998bc,Aharony:1999ti,Witten:1998zw}, a duality between a strongly coupled $\mathcal{N}=4$ Super Yang-Mills(SYM) theory and a classical supergravity in the asymptotic $AdS_5\times S^5$ background in the limit of large $N_c$ and strong t'Hooft coupling, is thus introduced to analyze the strongly coupled QCD. Although the degrees of freedom and properties of SYM theory are drastically distinct from those of QCD, the gauge/gravity duality could be an useful tool for providing a qualitative understanding of QGP and other physics in relativistic heavy ion collisions. The study of photon and dilepton production via the AdS/CFT correspondence was initiated by \cite{CaronHuot:2006te}. Then the influence of the inclusion of fundamental matters, which corresponds to massive quarks in the medium, on the photon production was studied in \cite{Mateos:2007yp}. In addition, the computations in Sakai-Sugimoto model \cite{Sakai:2004cn} as a QCD gravity dual with finite chemical potential and in SYM dual with finite density can be found in \cite{Parnachev:2006ev} and \cite{Jo:2010sg}, respectively. The study of photoemission rate and conductivity in multiple QCD duals is presented in \cite{PhysRevD.86.026003}. Recently, the photon spectrum
at intermediate coupling in SYM has been investigated as well \cite{Hassanain:2011ce}. Moreover, the studies of prompt photons and dileptons created in early times are presented in \cite{Baier:2012tc,Baier:2012ax,Steineder:2012si,Steineder:2013ana}.  

Nonetheless, a new mechanism beyond the isotropic production of thermal photons in QGP has to be introduced to trigger the excess of photon flow. In the weakly coupled scenario, different causes have been proposed. In \cite{vanHees:2011vb}, it is indicated that the considerable hadronic flow should enhance the photon $v_2$. On the other hand, the influence of pressure anisotropy of the medium on photon production has been discussed in \cite{Schenke:2006yp}. Also, the enhancement of photon production in the presence of strong magnetic field generated by two colliding nuclei in the early stage has been studied from a variety of approaches \cite{Tuchin:2010gx,Tuchin:2012mf,Basar:2012bp,Fukushima:2012fg,Bzdak:2012fr}.
In the strongly coupled scenario, the electromagnetic signatures in anisotropic plasma have been studied in \cite{Rebhan:2011ke,Patino:2012py}. In \cite{Patino:2012py}, an anisotropic dual geometry induced by an axion-dilaton gravity action introduced by Mateos and
Trancanelli(MT) \cite{Mateos:2011ix,Mateos:2011tv} is employed to mimic an anisotropic and thermalized plasma. It is found in \cite{Patino:2012py} that the photon production along the anisotropic direction with larger pressure increases as the anisotropy increases. Furthermore, the photon production from magnetic field in holography has been recently investigated by using the D4/D6 and D3/D7 systems in \cite{PhysRevD.87.026005}, where   
the photon production perpendicular to the magnetic field is found to be enhanced \cite{PhysRevD.87.026005}. Also, the elliptic flow $v_2$ and the coefficient of higher-order Fourier harmonic $v_4$ are computed in Sakai-Sugimoto model with magnetic field \cite{Yee:2013qma}.

Although the authors in \cite{Patino:2012py, PhysRevD.87.026005} evaluate the photon spectra with fundamental matters by applying D3/D7 embeddings, only the trivial embeddings, which correspond to contributions from massless quarks, are considered therein. As pointed out in \cite{Mateos:2007yp}, the contributions from massive quarks should be relatively considerable. To add fundamental matters to the background dominated by adjoint matters, we have to embed flavor probe branes in the background geometry \cite{Karch:2000gx,Karch:2002sh}. In the D3/D7 model, the probe D7 branes share the spacetime dimensions with the D3 branes, which generate the background geometry, and extend along the fifth dimension in the bulk; they further wrap an $S^3$ inside the $S^5$. The radius of $S^3$ will shrink to zero at the end of D7 branes in the bulk and the quark mass is associated with where the D7 branes end. At finite temperature, the D7 branes will be attracted by the black D3 branes. At low temperature or large quark mass, the D7 branes can fully remain outside the black branes. Such embeddings are referred to as the Minkowski embeddings corresponding to the confinement of mesons. On the other hand, at high temperature or small quark mass, the D7 branes will partially reside in the black branes. The embeddings are referred to as the black hole embeddings corresponding to meson melting in thermalized media. There exists a critical embedding when the D7 branes end at the event horizon, where the first order phase transition occurs\cite{Mateos:2011tv}. In our work, we will only consider black hole embeddings as deconfined phase.       

In this paper, we will investigate the thermal photon production with constant magnetic field and massive quarks in a strongly coupled, thermalized, and anisotropic plasma. In the holographic dual, we consider black hole embeddings coupled to constant magnetic field in the MT geometry, which can be regarded as generalization and extension of the previous studies in \cite{Patino:2012py,PhysRevD.87.026005}. The paper is organized in the following order. In section II, we briefly review the AdS/CFT prescription for the computations of spectral functions and evaluate the spectra of thermal photons in the MT metric with adjoint matters. In section III, we then consider the black hole embeddings in the MT geometry and compute the photon spectra with massive quarks. In section IV, we further incorporate constant magnetic field and study the photon spectra and DC conductivity in the presence of both magnetic field and pressure anisotropy. Finally, we make detailed discussions of our results in the last section.     
  
\section{\label{sec:level1}Spectral Functions in an Anisotropic Plasma}
To study the production of photons and dileptons from a strongly coupled and anisotropic plasma, we will evaluate spectral functions via the AdS/CFT correspondence in the dual geometry led by a five-dimensional dilaton-axion gravity action \cite{Mateos:2011ix,Mateos:2011tv}. In the Einstein frame, the action takes the form,
\begin{eqnarray}\label{MTaction}
S_E=\frac{1}{2\kappa^2}\int d^5x\sqrt{-g}\left(R+12-\frac{1}{2}(\partial\phi)^2-\frac{1}{2}e^{2\phi}(\partial\chi)^2\right)
+\frac{1}{2\kappa^2}\int d^4x\sqrt{-\gamma}2K,
\end{eqnarray}
where $\phi$ and $\chi$ denote the dilaton and axion, respectively. The second term in (\ref{MTaction}) is the Gibbons-Hawking-York boundary term and $2\kappa^2=16\pi G=8\pi^2/N_c^2$ is the five dimensional gravitational coupling. To simplify the computations, we have set $L=1$, where $L=(4\pi g_sN_cl_s^2)^{1/4}$ denotes the radius of $S^5$ in the ten-dimensional spacetime.
The solution of the dual metric in the Einstein frame is given by
\begin{eqnarray}\label{MT}
ds_E^2=\frac{e^{-\frac{\phi(u)}{2}}}{u^2}\left(-\mathcal{F}(u)\mathcal{B}(u)dt^2+dx^2+dy^2+\mathcal{H}(u)dz^2+\frac{du^2}{\mathcal{F}(u)}\right)
\end{eqnarray} 
for $\chi=az$, where $\mathcal{F}(u)$, $\mathcal{B}(u)$, and $\mathcal{H}(u)=e^{-\phi(u)}$ depend on $\phi(u)$ and the anisotropic factor $a$, which corresponds to the density of D7 branes embedded along the anisotropic direction $z$. These D7 branes dissolve in the bulk and contribute to the pressure anisotropy of the medium. The blackening function $\mathcal{F}(u)$ vanishes at the event horizon $u=u_h$, which results in the temperature and entropy density of the plasma. From \cite{Mateos:2011ix,Mateos:2011tv}, the temperature and entropy density of the MT geometry read
\begin{eqnarray}\nonumber
T&=&\sqrt{\mathcal{B}(u_h)}\frac{e^{\frac{1}{2}(\tilde{\phi}_{b}-\tilde{\phi}_h)}}{16\pi z_h}(16+u_h^2e^{\frac{7}{2}\tilde{\phi}_h}),\\
s&=&\frac{N_c^2a^{\frac{5}{7}}e^{-\frac{5}{4}\tilde{\phi}_h}}{2\pi u_h^3},
\end{eqnarray} 
where $\tilde{\phi}(z)=\phi(z)+\log a^{4/7}$, $\tilde{\phi}_b=\tilde{\phi}(0)$ and $\tilde{\phi}_h=\tilde{\phi}(u_h)$. At mid-anisotropy for $a/T=4.4$ or equivalently $a/s^{1/3}=1.2$, the medium forms a prolate in the momentum space with the ratio of pressures $P_z/P_{x,y}\approx 1.5$. In the rest of the paper, we will focus on the mid-anisotropy region when considering the anisotropic effect.  

To investigate the spectra of photons and dileptons, we will follow the approaches in \cite{CaronHuot:2006te} by introducing $U(1)$ gauge fields in the gravity dual as sources of electromagnetic currents on the boundary, where the gauge fields here are regarded as external probes and their back-reaction to the dual geometry is neglected. The effective action for the external gauge fields can be written as
\begin{eqnarray}\label{Sext}
S_{\mbox{ext}}=\frac{-1}{8\kappa^2}\int d^5x\sqrt{-g}F_{MN}F^{MN},
\end{eqnarray}
which leads to field equations $\nabla_{M}F^{MN}=0$, where the Latin indices denote the directions of five-dimensional spacetime, $M, N=t$, $x$, $y$, $z$, $u$. We will then choose the gauge $A_u=0$. Based on the translational invariance along $t$, $x$, $y$, $z$ directions, we can write down the Fourier transform of gauge fields as 
\begin{eqnarray}\label{FTforA}
A_{\mu}(u,t,\vec{x})=\int\frac{d^4k}{(2\pi)^4}e^{ik_0 t+i\vec{k}\cdot\vec{x}}A_{\mu}(u,k),
\end{eqnarray}
where the Greek indices denote the directions of four dimensional spacetime, $\mu=t$, $x$, $y$, $z$. Since now the dual geometry is anisotropic along the $z$ direction and rotational symmetry is only preserved on the $x-y$ plane, we will study two particular cases for $k=(-\omega,0,0,q)$ and $k=(-\omega,q,0,0)$, where the induced currents move parallel and perpendicular to the anisotropic direction, respectively. 

We firstly consider the case for $k=(-\omega,0,0,q)$; the field equations are
\begin{eqnarray}\label{eom1}\nonumber
A_{\perp}''&+&\left(\frac{\mathcal{F}'}{\mathcal{F}}-\frac{1}{u}+\frac{\mathcal{B}'}{2\mathcal{B}}+\frac{\mathcal{H}'}{2\mathcal{H}}-\frac{\phi'}{4}\right)A_{\perp}'
+\frac{1}{\mathcal{F}^2}\left(\frac{\omega^2}{\mathcal{B}}-\frac{q^2\mathcal{F}}{\mathcal{H}}\right)A_{\perp}=0,\\
\nonumber A_t''&-&\left(\frac{1}{u}+\frac{\mathcal{B}'}{2\mathcal{B}}-\frac{\mathcal{H}'}{2\mathcal{H}}+\frac{\phi'}{4}\right)A_t'-\frac{q}{\mathcal{F}\mathcal{H}}(qA_t+\omega A_z)=0,\\
\nonumber
A_z''&+&\left(\frac{\mathcal{F}'}{\mathcal{F}}-\frac{1}{u}+\frac{\mathcal{B}'}{2\mathcal{B}}-\frac{\mathcal{H}'}{2\mathcal{H}}-\frac{\phi'}{4}\right)A_z'+\frac{\omega}{\mathcal{F}^2\mathcal{B}}
(\omega A_z+qA_t)=0,
\\
A_t'&=&-\frac{q\mathcal{B}\mathcal{F}}{\omega\mathcal{H}}A_z',
\end{eqnarray}
where $A_{\mu}=A_{\mu}(u,k)$ and primes denote the derivatives with respect to $u$. The first equation in (\ref{eom1}) for $\perp=x,y$ is the field equation for transverse polarizations perpendicular to the spatial momentum. The rest three equations govern the longitudinal modes, where the last one is in fact redundant, which can be obtained from the linear combination of the other two equations. By defining $E_{\perp}=\omega A_{\perp}$ and $E_z=qA_t+\omega A_z$, we can rewrite the field equations into gauge invariant forms,
\begin{eqnarray}\label{EOM1}
\nonumber
&&E_{\perp}''+\left(\frac{\mathcal{F}'}{\mathcal{F}}-\frac{1}{u}+\frac{\mathcal{B}'}{2\mathcal{B}}+\frac{\mathcal{H}'}{2\mathcal{H}}-\frac{\phi'}{4}\right)E_{\perp}'
+\frac{1}{\mathcal{F}^2}\left(\frac{\omega^2}{\mathcal{B}}-\frac{q^2\mathcal{F}}{\mathcal{H}}\right)E_{\perp}=0,\\
&&E_z''+\left(\frac{\omega^2\mathcal{H}\mathcal{F}'}{\mathcal{F}}+\left(\frac{\mathcal{B}'}{2\mathcal{B}}-\frac{\mathcal{H}'}{2\mathcal{H}}\right)(\omega^2\mathcal{H}+q^2\mathcal{B}\mathcal{F})\right)
\frac{E_z'}{\omega^2\mathcal{H}-q^2\mathcal{B}\mathcal{F}}-\left(\frac{1}{u}+\frac{\phi'}{4}\right)E_z'\nonumber\\
&&+\frac{1}{\mathcal{F}^2}\left(\frac{\omega^2}{\mathcal{B}}-\frac{q^2\mathcal{F}}{\mathcal{H}}\right)E_z=0,
\end{eqnarray}
for $k=(-\omega,0,0,q)$. Similarly, we have
\begin{eqnarray}\label{EOM2}
\nonumber
&&E_y''+\left(\frac{\mathcal{F}'}{\mathcal{F}}-\frac{1}{u}+\frac{\mathcal{B}'}{2\mathcal{B}}+\frac{\mathcal{H}'}{2\mathcal{H}}-\frac{\phi'}{4}\right)E_y'
+\frac{1}{\mathcal{F}^2}\left(\frac{\omega^2}{\mathcal{B}}-q^2\mathcal{F}\right)E_y=0,\\
\nonumber
&&E_z''+\left(\frac{\mathcal{F}'}{\mathcal{F}}-\frac{1}{u}+\frac{\mathcal{B}'}{2\mathcal{B}}-\frac{\mathcal{H}'}{2\mathcal{H}}-\frac{\phi'}{4}\right)E_z'
+\frac{1}{\mathcal{F}^2}\left(\frac{\omega^2}{\mathcal{B}}-q^2\mathcal{F}\right)E_z=0,\\
&&E_x''+\left(\frac{\omega^2\mathcal{F}'}{\mathcal{F}}+\frac{\mathcal{B}'}{2\mathcal{B}}(\omega^2+q^2\mathcal{B}\mathcal{F})\right)
\frac{E_x'}{\omega^2-q^2\mathcal{B}\mathcal{F}}-\left(\frac{1}{u}+\frac{\phi'}{4}-\frac{\mathcal{H}'}{2\mathcal{H}}\right)E_x'\nonumber\\
&&+\frac{1}{\mathcal{F}^2}\left(\frac{\omega^2}{\mathcal{B}}-q^2\mathcal{F}\right)E_x=0,
\end{eqnarray}
for $k=(-\omega,q,0,0)$, where $E_{y,z}=\omega A_{y,z}$ and $E_x=qA_t+\omega A_x$. Notice that the equation for $E_y$ here is slightly different from that for $E_z$ due to anisotropy of the metric along the $z$ direction. By using the field equations (\ref{EOM1}) and (\ref{EOM2}), we then write down the boundary terms of the effective action in (\ref{Sext}),
\begin{eqnarray}
S_{\epsilon}&=&\frac{-1}{4\kappa^2}\int \frac{d^4k}{(2\pi)^4}\frac{Q}{u}\left(\frac{1}{\omega^2}E^*_{\perp}E'_{\perp}+\frac{1}{\omega^2\mathcal{H}-q^2\mathcal{B}\mathcal{F}}
E^*_zE_z'\right),\mbox{ for }k=(-\omega,0,0,q),\\\nonumber
S_{\epsilon}&=&\frac{-1}{4\kappa^2}\int \frac{d^4k}{(2\pi)^4}\frac{Q}{u}\left(\frac{1}{\omega^2}(E^*_{y}E'_{y}+\mathcal{H}^{-1}E^*_{z}E'_{z})+\frac{1}{\omega^2-q^2\mathcal{B}\mathcal{F}}
E^*_xE_x'\right),\mbox{ for }k=(-\omega,q,0,0),
\end{eqnarray}
where $E^*_{\mu}=E(u,-k)$ and $Q=\mathcal{F}\sqrt{\mathcal{B}}e^{-\frac{3}{4}\phi}$. Notice that $\phi\rightarrow 0$ and $\mathcal{F}, \mathcal{B}, \mathcal{H}\rightarrow 1$ near the boundary. Although the dual geometry asymptotically reduces to the pure AdS spacetime near the boundary, the boundary value of $E_y$ and of $E_z$ will differ due to the breaking of rotational symmetry in the bulk.
  
In thermal equilibrium, the differential photon emission rate per unit volume can be written as 
\begin{eqnarray}\label{dGamma}
d\Gamma_{\gamma}=
\frac{d^3k}{2(2\pi)^3}\frac{\chi(k)}{\omega(e^{\beta \omega}-1)},\quad \chi(k)
=-2\mbox{Im}[\sum^n_{s=1}\epsilon_s^{\mu}\epsilon_s^{*\nu}C_{\nu\mu}(k)],
\end{eqnarray}
where $n=2$ denotes the number of polarizations of photons and $\chi(k)$ represents the trace of the spectral density, which is related to the retarded current-current correlator $C_{\mu\nu}(k)$. When photons are linearly polarized along a particular polarization $\epsilon_T$, we should take
\begin{eqnarray}\label{chiC}
d\Gamma_{\gamma}(\epsilon_T)=
\frac{d^3k}{2(2\pi)^3}\frac{\chi_{\epsilon_T}(k_0)}{\omega(e^{\beta \omega}-1)},\quad \chi_{\epsilon_T}(k_0)=-4\mbox{Im}[\epsilon_T^{\mu}\epsilon_T^{*\nu}C_{\nu\mu}(k)].
\end{eqnarray}
By these definitions, we retrieve $\chi_{\epsilon_T}(k_0)=\chi(k)$ in the isotropy case. 
Following the AdS/CFT prescription\cite{Son:2002sd,Kovtun:2005ev,CaronHuot:2006te}, the retarded correlators can be evaluated by taking the functional derivatives of the boundary action with respect to the gauge fields, which further results in 
\begin{eqnarray}\label{eqPi}
\chi(k)=\mathcal{Z}\mbox{Im}\lim_{u\rightarrow 0}\frac{QE'_{\perp}(u,k)}{uE_{\perp}(u,k)},\quad\chi_{\epsilon_T}(k_0)=\mathcal{Z}\mbox{Im}\lim_{u\rightarrow 0}\frac{QE'_{\epsilon_T}(u,k)}{uE_{\epsilon_T}(u,k)},
\end{eqnarray}
where $\mathcal{Z}$ is an overall constant. Also, the zero-frequency limit of the spectral function contributes to the DC conductivity as
\begin{eqnarray}\label{Dcconductivity}
\sigma=\frac{e^2}{4}\lim_{\omega\rightarrow 0}\frac{1}{\omega}\chi(k)|_{|\vec{k}|=\omega},\quad \sigma(\epsilon_T)=\frac{e^2}{4}\lim_{\omega\rightarrow 0}\frac{1}{\omega}\chi_{\epsilon_T}(k_0)|_{|\vec{k}|=\omega}.
\end{eqnarray} 
In thermal equilibrium, only the incoming-wave solutions near the horizon have to be considered. In the isotropic case, the lightlike solution($q=\omega$) takes the form \cite{CaronHuot:2006te},
\begin{eqnarray}
E^{T}_{in}(\omega,u)=\left(1-\frac{u^2}{u_h^2}\right)^{-\frac{i\hat{\omega}}{4}}\left(1+\frac{u^2}{u_h^2}\right)^{-\frac{\hat{\omega}}{4}}
{}_2F_1\left(1-\frac{1+i}{4}{\hat{\omega}},-\frac{1+i}{4}{\hat{\omega}};
1-i\frac{\hat{\omega}}{2};\frac{(1-u^2/u_h^2)}{2}\right),
\end{eqnarray}
where $u_h=(\pi T)^{-1}$ in the isotropic geometry and $\hat{\omega}=\omega/(\pi T)$. However, the anisotropic solutions can only be solved numerically. We thus solve $E_{\perp}$ in (\ref{EOM1}) and $E_y$, $E_z$ in (\ref{EOM2}) for $q=\omega$ by imposing incoming-wave boundary conditions via analysis of the near-horizon expansion of field equations. In the first case,
by solving $E_{\perp}$ and implementing (\ref{eqPi}), we can derive the spectral density for photons propagating along the anisotropic direction $z$. In the second case, the solutions of $E_y$ and $E_z$ then contribute to the spectra for photons moving perpendicular to the anisotropic direction. 

\begin{figure}[h]
\begin{center}
{\includegraphics[width=7.5cm,height=5cm,clip]{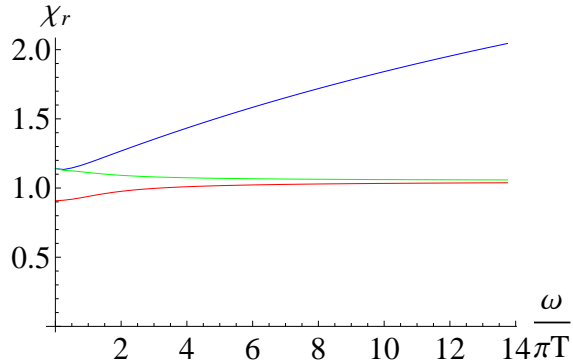}}
\caption{The blue, green, and red curves(from top to bottom) represent the ratios of spectral densities at fixed temperature for $\epsilon_T=\epsilon_x$ or $\epsilon_y$ when $k=(-\omega,0,0,\omega)$, for $\epsilon_T=\epsilon_y$ and $\epsilon_T=\epsilon_z$ when $k=(-\omega,\omega,0,0)$, respectively. Here we take $u_h=1$ and $a/T=4.4$.}\label{chir}
\end{center}
\end{figure}

To compare the results with those found in isotropic case, we have to fix certain physical scales such as temperature or entropy density of the media. In the rest of the paper, we will solely focus on the cases with fixed temperature. 
Due to the rotational symmetry on the $x-y$ plane, the computation of the spectral density for $k=(-\omega,0,0,\omega)$ is straightforward. We define the ratio $\chi_r=\chi_{\epsilon_T\mbox{aniso}}/\chi_{\mbox{iso}}$ to compare the anisotropic and isotropic cases. Nonetheless, as indicated in the previous context, the field equations for $E_y$ and $E_z$ are slightly different in (\ref{EOM2}) for $k=(-\omega,\omega,0,0)$, which bring about different retarded correlators \footnote[1]{Here $\chi_{r}$ is equivalent to $2\chi_{(1,2)}/\chi_{iso}(T)$ defined in \cite{Patino:2012py}. Also,  $k=(-\omega,0,0,\omega)$ and $k=(-\omega,\omega,0,0)$ correspond to $\theta=0$ and $\theta=\pi/2$ therein.}. 
The results at mid anisotropy for fixed temperature are shown in Fig.\ref{chir}, which match those found by fluctuating a flavor probe brane in MT geometry in the massless-quark limit \cite{Patino:2012py}\footnote{Following the convention therein, $\chi_{\epsilon_T}=2\chi_{(1(2))}$.}.
The matching is expected since only the leading-order contribution of the gauge fields coupled to the flavor brane in the DBI action is considered in \cite{Patino:2012py}.

\section{\label{sec:level1}Embedded Flavor Branes}
To incorporate the flavor degrees of freedom, we will consider the embeddings of flavor D7 branes \cite{Karch:2000gx,Karch:2002sh}, which are different from the D7 branes generating pressure anisotropy of the background geometry \cite{Mateos:2011ix,Mateos:2011tv}. In addition, we take the quenched approximation by assuming $N_f\ll N_c$, where $N_f$ denotes the number of flavors. With such an approximation, the modification of flavor probe branes to the background geometry is $\mathcal{O}(N_f/N_c)$ suppressed, which will be neglected in this paper. The flavor D7 brane extended into the bulk shares the same spacetime dimensions with the D3 brane on the boundary and wraps an $S^3$ inside the $S^5$,
\begin{eqnarray}
d\Omega^2_5=d\theta^2+\sin^2\theta d\Omega^2_3+\cos^2\theta d\eta^2.
\end{eqnarray}
For convenience, we hereafter work in the string frame. The flavor D7 brane is characterized by the Dirac-Born-Infeld (DBI) action,
\begin{eqnarray}\label{DBI}
S=-N_fT_{D7}\int_{D7} d^8x e^{-\phi}\sqrt{-\mbox{det}(G+2\pi l_s^2F)},
\end{eqnarray}
where $G$ is the induced metric on the D7 branes, $F=dA$ is the U(1) field strength for the massless gauge fields coupled to the brane, and $T_{D7}=(2\pi l_s)^{-7}(g_sl_s)^{-1}$ is the D7-brane string tension. The induced metric in the string frame led by the embedding of the flavor branes in the MT geometry reads \cite{Patino:2012py}
\begin{eqnarray}\label{D7brane}\nonumber
ds^2_{D7}&=&\frac{1}{u^2}\left(-\mathcal{F}(u)\mathcal{B}(u)dt^2+dx^2+dy^2+\mathcal{H}(u)dz^2\right)+\frac{1-\psi(u)^2+u^2\mathcal{F}(u)e^{\frac{1}{2}\phi(u)}\psi(u)'^2}{u^2\mathcal{F}(u)(1-\psi(u)^2)}du^2\\
&&+e^{\frac{1}{2}\phi(u)}(1-\psi(u)^2)d\Omega_3^2,
\end{eqnarray}
where $\sqrt{1-\psi(u)^2}=\sin\theta$ represents the radius of the internal $S^3$ wrapped by the D7 branes. To compute the leading-order contribution of photon spectra, we could preserve the quadratic order in the field strength for the D7-brane action,
\begin{eqnarray}\label{DBIexp}
S=-N_fT_{D7}\int_{D7} d^8x e^{-\phi}\sqrt{-\mbox{det}(G)}\left(1+\frac{(2\pi l_s)^2}{4}F^2\right).
\end{eqnarray}
By treating the expansion of the field strength in the DBI action perturbatively, we could neglect the back-reaction of gauge fields to the induced metric. The D7-brane action now reads
\begin{eqnarray}\label{DBIaniso}
S=-K_{D7}\int dtd^3\vec{x}duF^2\frac{(1-\psi^2)e^{-\frac{3\phi}{4}}}{u^5}
\sqrt{\mathcal{B}(1-\psi^2+u^2\mathcal{F}e^{\frac{1}{2}\phi}\psi^{\prime 2})},
\end{eqnarray}
where the prefactor $K_{D7}$ includes the integration over the internal space $\Omega_3$ wrapped by the D7 branes.
By following the standard AdS/CFT prescription as shown in the previous section, the spectral functions of electromagnetic probes can be obtained. In the case of trivial embedding $\psi=0$, which corresponds to the massless-quark case, the spectra reduce to the results we acquired in the previous section except for the difference in the prefactors.

To incorporate massive quarks, we have to consider nontrivial embeddings $\psi\neq 0$. In the absence of the gauge fields, we derive the field equation of $\psi$ by minimizing the action in (\ref{DBIexp}),
\begin{eqnarray}\label{psiEOM}\nonumber
\psi''(u)&+&\mathcal{C}_1\psi'(u)+\mathcal{C}_2\psi'(u)^3+\mathcal{C}_3\psi(u)+\mathcal{C}_4\psi(u)^3=0,\\
\nonumber
\mathcal{C}_1&=&\frac{\mathcal{B}'}{2\mathcal{B}}+\frac{\mathcal{F}'}{\mathcal{F}}+\frac{\mathcal{H}'}{2\mathcal{H}}-\frac{3}{u}+\frac{\phi'}{4},\\
\nonumber
\mathcal{C}_2&=&-\frac{4e^{\frac{\phi}{2}}\mathcal{F}u}{(1-\psi^2)}+\frac{e^{\frac{\phi}{2}}\mathcal{F}u^2}{2(1-\psi^2)}\left(\frac{\mathcal{F}'}{\mathcal{F}}+\frac{\mathcal{B}'}{\mathcal{B}}+\frac{\mathcal{H}'}{\mathcal{H}}\right),\\
\nonumber
\mathcal{C}_3&=&\frac{3e^{\frac{-\phi}{2}}}{u^2\mathcal{F}(1-\psi^2)}+\frac{4\psi'^2}{(1-\psi^2)},\\
\mathcal{C}_4&=&-\frac{3e^{\frac{-\phi}{2}}}{u^2\mathcal{F}(1-\psi^2)}.
\end{eqnarray}

The field equation in (\ref{psiEOM}) can be solved by imposing proper boundary conditions near the horizon \cite{Mateos:2006nu,Hoyos:2006gb}, where we take $\psi(u_h)=\psi_0$ and $\psi'(u_h)=(-3u^{-2}e^{-\phi/2}\psi/\mathcal{F}')|_{u=u_h}$. Here we only consider the black hole embedding as a deconfined phase of the plasma, which corresponds to the choice of $0\leq\psi_0<1$ \cite{Mateos:2006nu}.

In the isotropic case, the asymptotic solution of $\psi(u)$ near the boundary behaves as
\begin{eqnarray}
\psi(u)=m\frac{u}{2^{1/2}u_h}+c\frac{u^3}{2^{3/2}u_h^3}+\dots,
\end{eqnarray}
where the dimensionless coefficients $m$ and $c$ are related to the magnitudes of quark mass and condensate through \cite{Mateos:2006nu,Mateos:2007vn}
\begin{eqnarray}\nonumber
M_q&=&\frac{m}{2^{3/2}\pi l_s^2u_h}=\frac{\sqrt{\lambda}m}{2\pi u_h},\\
\langle\mathcal{O}\rangle&=&-2^{3/2}\pi^3l_s^2N_fT_{D7}u_h^{-3}c=
-\frac{\sqrt{\lambda}N_cN_f}{8\pi^3u_h^3}c.
\end{eqnarray}
Recall that $l_s^2=L^2/(2\sqrt{\pi g_sN_c})=1/\sqrt{2\lambda}$ for $\lambda=g^2_{YM}N_c=2\pi g_sN_c$ being the t'Hooft coupling and $L=1$ in our convention.
However, at nonzero anisotropy, the asymptotic solution will contain extra logarithmic terms, 
\begin{eqnarray}
\psi(u)=m\frac{u}{2^{1/2}u_h}+c\frac{u^3}{2^{3/2}u_h^3}+a^2\rho_3\frac{u^3}{u_h^3}\log(u/u_h)\dots,
\end{eqnarray}   
which come from the anisotropy. The numerical computations of $c$ may become technically difficult due to the presence of the leading logarithmic term, while the extraction of $m$ is straightforward. Further discussions of black hole embeddings in the MT geometry can be found in Appendix B.   
In Fig.\ref{psi0Bm}, the quark mass scaled by temperature with respect to $\psi_0$ for the black hole embeddings with anisotropy is represented by the dashed blue curve, while the solid blue curve corresponds to the isotropic case. The critical mass at $\psi_0\rightarrow 1$ now is increased by anisotropy or equivalently the dissociation temperature is reduced. In fact, as shown in \cite{Mateos:2007vn}, the black hole embedding near $\psi_0=1$ could be metastable or unstable and the phase transition occurs within this region in the isotropic case. To manifest the phase transition near $\psi_0=1$ in the anisotropic case requires further investigation on the thermodynamics of the flavor brane in Minkowski embeddings, which is beyond the scope of this paper. The relevant work is in progress \cite{wuyang}.

After solving the induced metric of the flavor probe brane, we can now compute photon spectra. By taking the Fourier transform of gauge fields as (\ref{DBIaniso}), the action near the boundary becomes
\begin{eqnarray}\label{Saniso}
S_{\epsilon}&=&-2K_{D7}\int \frac{d^4k}{(2\pi)^4}\frac{Q_{D7}}{u}\left(-\frac{1}{\mathcal{F}\mathcal{B}}A^*_tA_t'+
A^*_{\perp}A_{\perp}'+\frac{1}{\mathcal{H}}A^*_zA_z'\right),\\
Q_{D7}&=&\nonumber\frac{(1-\psi^2)^2\sqrt{\mathcal{B}}\mathcal{F}e^{-\frac{3\phi}{4}}}
{\sqrt{1-\psi^2+u^2\mathcal{F}e^{\frac{1}{2}\phi}\psi^{\prime 2}}},
\end{eqnarray}
where the gauge fields have to obey the field equations
\begin{eqnarray}\label{anisoeom}
\partial_{\mu}(M G^{\mu\alpha}G^{\nu\beta}F_{\alpha\beta})=0,\quad M=\frac{(1-\psi^2)e^{-\frac{3\phi}{4}}}{u^5}
\sqrt{\mathcal{B}(1-\psi^2+u^2\mathcal{F}e^{\frac{1}{2}\phi}\psi^{\prime 2})}.
\end{eqnarray}
Recall that $G_{\mu\nu}$ here is the induced metric of D7 branes. To convert (\ref{Saniso}) and (\ref{anisoeom}) into gauge-invariant forms, we may follow the general derivation presented in Appendix A.
From (\ref{gchi}) and $\mathcal{H}\rightarrow 1$ near the boundary, we have the photon spectral density
\begin{eqnarray}
\chi_{\epsilon_j}(\omega)=8K_{D7}\mbox{Im}\lim_{u\rightarrow 0}\frac{Q_{D7}E'_j(u,\omega)}{uE_j(u,\omega)},
\end{eqnarray}   
where $E_j(u,\omega)=\omega A_j(u,k)|_{k_0=-\omega}$ for $j$ being the transverse polarization. As discussed in the previous section, the transverse-polarized gauge field with $k=(-\omega,0,0,\omega)$ preserves rotational symmetry on the $x-y$ plane, which is governed by just one equation of motion. Whereas for the field with $k=(-\omega,\omega,0,0)$, the two types of  transverse polarizations $A_y$ and $A_z$ should obey different field equations due to the presence of anisotropy along the $z$ direction. The computations of solving the field equations can be carried out via the similar approaches introduced in the previous section. 

\begin{figure}[h]
\begin{minipage}{7cm}
\begin{center}
{\includegraphics[width=7.5cm,height=5cm,clip]{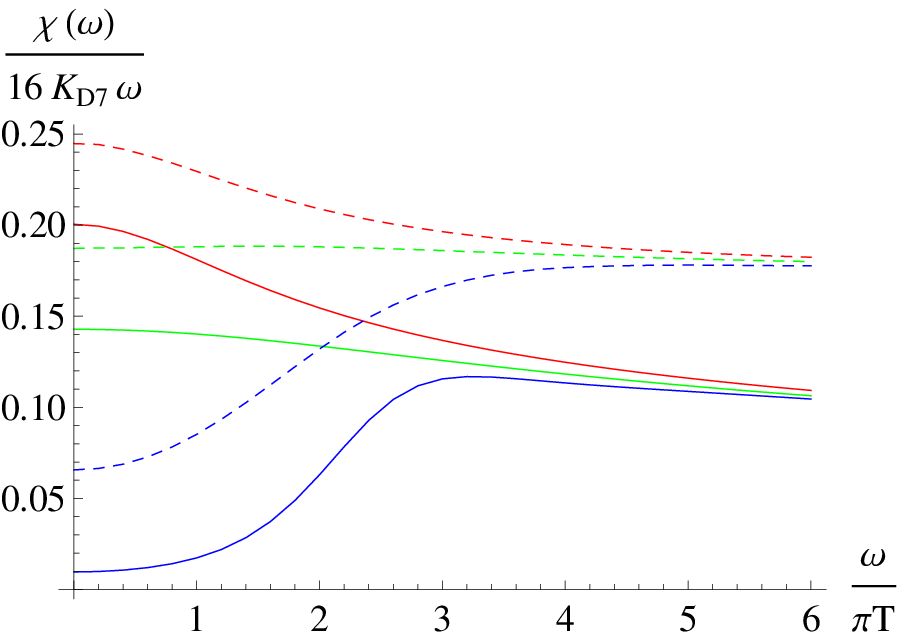}}
\caption{The red, green, and blue curves(from top to bottom) represent the spectral functions with $k=(-\omega,0,0,\omega)$ for $\hat{M}_q/(\pi T)=0.61$, $0.89$, and $1.31$.The dashed ones and solid ones correspond to the results with and without anisotropy, respectively. Here we take $u_h=1$ and $a/T=4.4$.}\label{anisospecqz}
\end{center}
\end{minipage}
\hspace {1cm}
\begin{minipage}{7cm}
\begin{center}
{\includegraphics[width=7.5cm,height=5cm,clip]{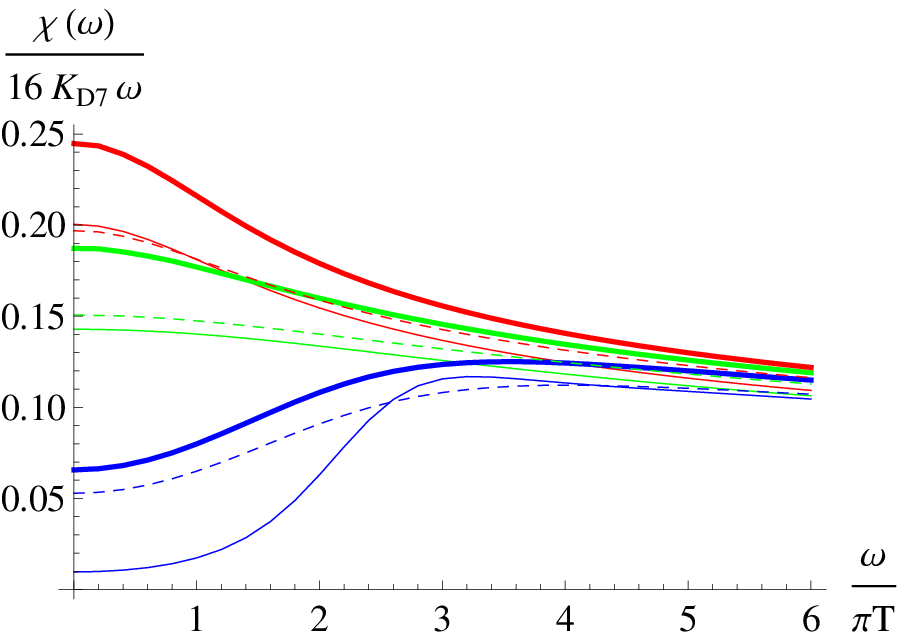}}
\caption{The red, green, and blue curves(from top to bottom) represent the spectral functions with $k=(-\omega,\omega,0,0)$ and the $y-$polarization for $\hat{M}_q/(\pi T)=0.61$, $0.89$, and $1.31$. The thin ones correspond to isotropic results. The thick ones and dashed ones correspond to the anisotropic results with $\epsilon_y$ and $\epsilon_z$, respectively. Here we take $u_h=1$ and $a/T=4.4$.}
\label{anisospecqx}
\end{center}
\end{minipage}
\end{figure}

To compare the anisotropic spectral densities with isotropic ones, we have to fix the quark mass and temperature of the media. Here we define the rescaled mass $\hat{M}_Q=2\pi M_q/\sqrt{\lambda}$. The spectral functions for photons moving along the anisotropic direction are shown in Fig.\ref{anisospecqz}, while the results for photons moving perpendicular to the anisotropic direction are illustrated in Fig.\ref{anisospecqx}, where the spectral functions are in the unit of $\pi T$. At small $\hat{\omega}=\omega/(\pi T)$, the spectral functions contributed from the quarks with different mass possess distinct features, while the qualitative structures of isotropic and anisotropic spectra are similar. At large $\hat{\omega}=\omega/(\pi T)$, the effect from the difference of quark mass are suppressed by the energy of photons; the spectra hence converge to the same amplitudes. Nevertheless, as shown in Fig.\ref{anisospecqz}, the anisotropic spectra for photons moving along the anisotropic direction receive overall enhancement in amplitudes. For the photons moving perpendicular to the anisotropic direction, the amplitudes of anisotropic spectra can be smaller or larger than the isotropic ones depending on the quark mass and the polarization as illustrated in Fig.\ref{anisospecqx}. At large $\hat{\omega}$, the anisotropic spectra for photons moving perpendicular to the anisotropic direction saturate the isotropic ones.  
 
\section{\label{sec:level1}External Magnetic Field}
\begin{figure}[h]
\begin{minipage}{7cm}
\begin{center}
{\includegraphics[width=7.5cm,height=5cm,clip]{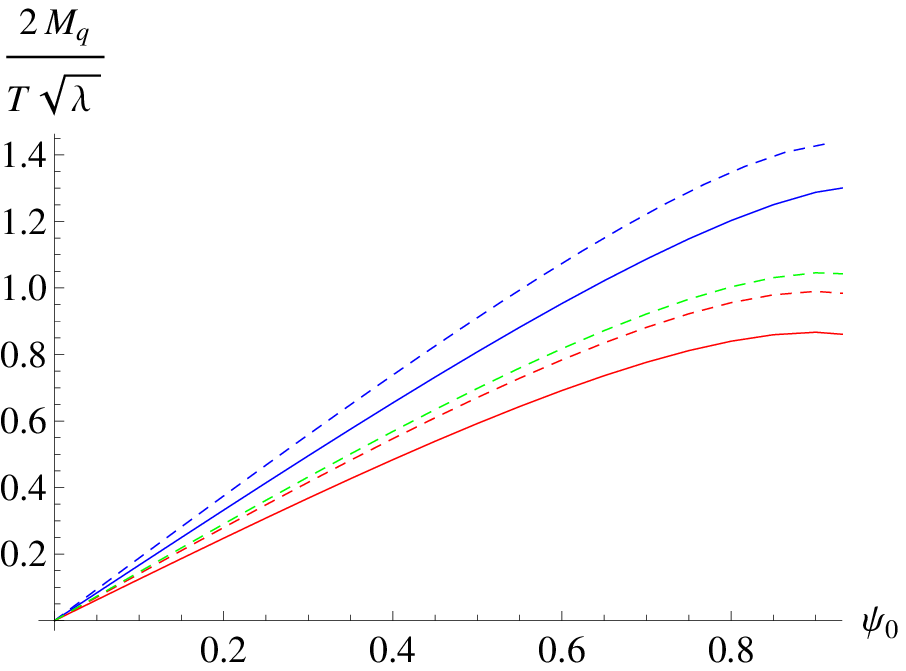}}
\caption{The blue and red solid curves(from top to bottom) represent the quark mass scaled by temperature without and with magnetic field $B_z$, respectively. The blue, green, and red dashed curves(from top to bottom) correspond to the anisotropic case without magnetic field, with $B_y$, and with $B_z$, respectively. Here we set $u_h=1$, $B_y=B_z=2(\pi T)^2$, and $a/T=4.4$.}\label{psi0Bm}
\end{center}
\end{minipage}
\hspace {1cm}
\begin{minipage}{7cm}
\begin{center}
{\includegraphics[width=7.5cm,height=5cm,clip]{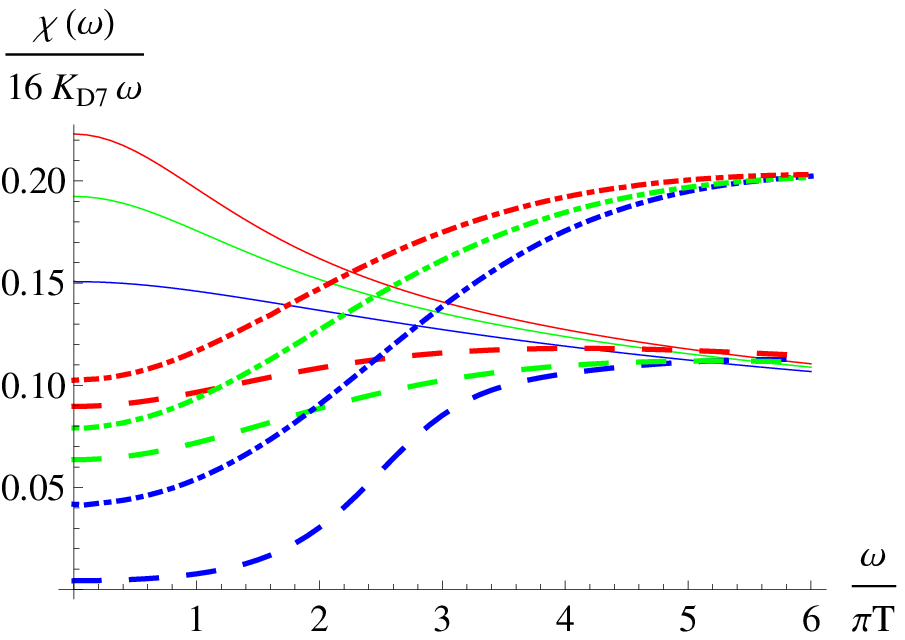}}
\caption{The red, green, and blue curves(from top to bottom) represent the spectral functions with $k=(-\omega,0,0,\omega)$ for $\hat{M}_q/(\pi T)=0.45$, $0.65$, and $0.86$, respectively. The solid, dashed, and dot-dashed correspond to $(a/T,B_z/(\pi T)^2)=(0,0)$, $(0,2)$, and $(4.4,2)$, respectively.}
\label{isoBzanisoBz}
\end{center}
\end{minipage}
\end{figure}

In D3/D7 embeddings, the flavor D7 branes can couple to external electromagnetic field via the DBI action\cite{Filev:2007gb,Albash:2007bk,Albash:2007bq}. Here we consider the presence of constant magnetic field by turning on the worldvolume U(1) gauge field $2\pi l_s^2A_y=B_z x$ and $2\pi l_s^2A_x=B_yz$ in (\ref{DBI}), which generate the magnetic field $B_z$ and $B_y$ along the $z$ and $y$ directions, respectively\footnote{Notice that the $B_{z(y)}$ here is the reparametrized magnetic field. The magnitude of the real magnetic field should be written as $eB=B_{z(y)}/(2\pi l_s^2)=B_{z(y)}\pi^{-1}\sqrt{\lambda/2}$ for $L=1$, which depends on the t'Hooft coupling.}. To simplify the computations, we will include the magnetic field in one of the directions alone. In the isotropic case, these two setups should degenerate, while the degeneracy will be broken when we further incorporate the pressure anisotropy. The explicit forms of the DBI actions now become 
\begin{eqnarray}\label{DBIanisoB}\nonumber
S&=&-N_fT_{D7}\int d^8x\frac{(1-\psi^2)e^{-\frac{3\phi}{4}}}{u^5}
\sqrt{\mathcal{B}(1+B_z^2u^4)(1-\psi^2+u^2\mathcal{F}e^{\frac{1}{2}\phi}\psi^{\prime 2})},\mbox{ and }\\
S&=&-N_fT_{D7}\int d^8x\frac{(1-\psi^2)e^{-\frac{\phi}{4}}}{u^5}
\sqrt{\mathcal{B}(\mathcal{H}+B_y^2u^4)(1-\psi^2+u^2\mathcal{F}e^{\frac{1}{2}\phi}\psi^{\prime 2})},
\end{eqnarray}
which lead to same field equations as (\ref{psiEOM}) with the following substitutions,
\begin{eqnarray}\nonumber
C_1\rightarrow C_1|_{B_z=0}+\frac{2B_z^2u^3}{1+B_z^2u^4},\qquad C_2\rightarrow C_2|_{B_z=0}+\frac{2 B_z^2 e^{\frac{\phi}{2}} u^5 \mathcal{F}}{\left(1+B_z^2 u^4\right) \left(1-\psi^2\right)},
\mbox{ and }\\
C_1\rightarrow C_1|_{B_y=0}+\frac{B_y^2 u^3 \left(4 \mathcal{H}-u \mathcal{H}'\right)}{2 \mathcal{H} \left(B_y^2 u^4+\mathcal{H}\right)},
\qquad C_2\rightarrow C_2|_{B_y=0}+\frac{B_y^2 e^{\frac{\phi}{2}} u^5 \mathcal{F} \left(4 \mathcal{H}-u \mathcal{H}'\right)}{2 \mathcal{H} \left(B_y^2 u^4+\mathcal{H}\right) \left(1-\psi^2\right)}.
\end{eqnarray}
The $\psi(u)$ in the field equation can be solved numerically with the same manner as in the cases with zero magnetic field. By analyzing the near-horizon expansions of $\psi$, one could find that the relation between $\psi_0$ and $\psi_0'$ is not altered by the inclusion of magnetic field. On the other hand, the leading order coefficients in the expansions of $\psi(u)$ near the boundary then contribute to quark mass. Different values of quark mass obtained by varying $\psi_0$ for black hole embeddings in the presence of magnetic field or anisotropy are illustrated in Fig.\ref{psi0Bm}. It is found that the magnetic field reduces the critical mass or increases the dissociation temperature, which results in an opposite effect to the pressure anisotropy. The suppression of critical mass by magnetic field has been found in \cite{Albash:2007bq} as well.

\begin{figure}[h]
\begin{minipage}{7cm}
\begin{center}
{\includegraphics[width=7.5cm,height=5cm,clip]{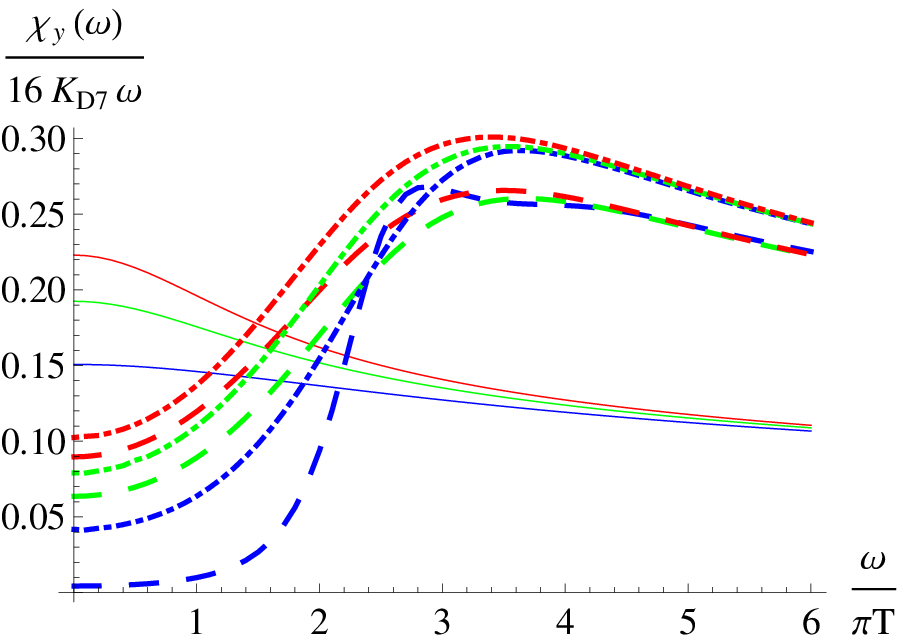}}
\caption{The red, green, and blue curves(from top to bottom) represent the spectral functions with $k=(-\omega,\omega,0,0)$ and $\epsilon_T=\epsilon_y$ for $\hat{M}_q/(\pi T)=0.45$, $0.65$, and $0.86$, respectively. The solid, dashed, and dot-dashed correspond to $(a/T,B_z/(\pi T)^2)=(0,0)$, $(0,2)$, and $(4.4,2)$, respectively.}\label{isoBzanisoBzqxAy}
\end{center}
\end{minipage}
\hspace {1cm}
\begin{minipage}{7cm}
\begin{center}
{\includegraphics[width=7.5cm,height=5cm,clip]{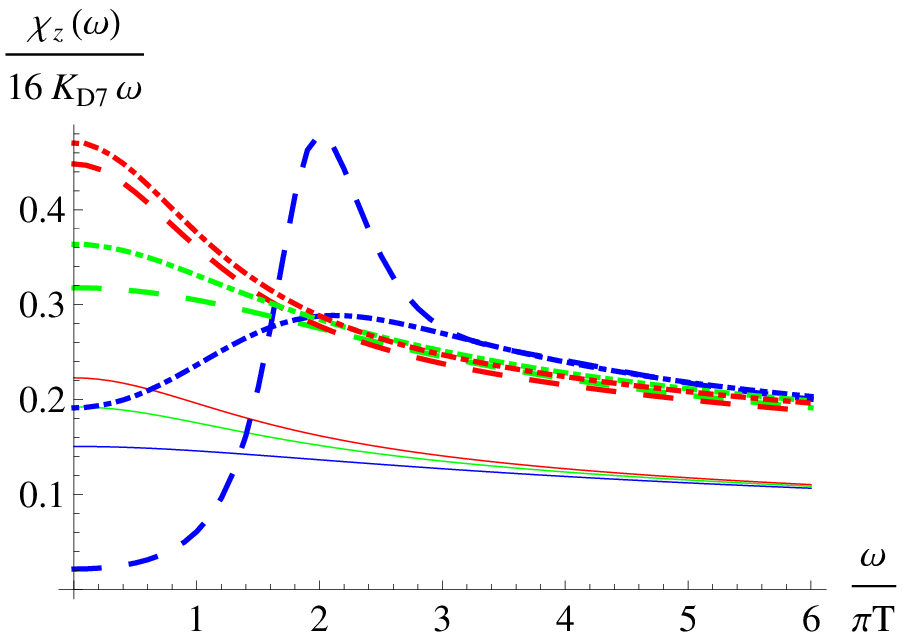}}
\caption{The red, green, and blue curves(from top to bottom) represent the spectral functions with $k=(-\omega,\omega,0,0)$ and $\epsilon_T=\epsilon_z$ for $\hat{M}_q/(\pi T)=0.45$, $0.65$, and $0.86$, respectively. The solid, dashed, and dot-dashed correspond to $(a/T,B_z/(\pi T)^2)=(0,0)$, $(0,2)$, and $(4.4,2)$, respectively.}
\label{isoBzanisoBzqxAz}
\end{center}
\end{minipage}
\end{figure}

To generate the electromagnetic currents on the boundary, we should further introduce the perturbation of gauge fields in the presence of magnetic field. To the quadratic order in the field strength, the D7-brane action is 
\begin{eqnarray}\label{DBIexpB}
S=-N_fT_{D7}\int_{D7} d^8x e^{-\phi}\sqrt{-\mbox{det}(G_{\mu\nu})}\left(1+\frac{(2\pi l_s)^2}{4}F^2\right),
\end{eqnarray}  
where $G_{\mu\nu}$ is now the induced metric of the D7 branes incorporating the magnetic field. The diagonal elements of $G_{\mu\nu}$ are the same as those in the absence of magnetic field, while the off-diagonal terms $G_{xy}(G_{xz})=-G_{yx}(G_{zx})=B_z(-B_y)$ receive the contributions from nonzero magnetic field. By taking Fourier transform of the gauge fields, the near boundary actions can be written as
\begin{eqnarray}\nonumber
S_{\epsilon}&=&-2K_{D7}\int \frac{d^4k}{(2\pi)^4}\frac{Q_{B_z}}{u}\left(-\frac{1}{\mathcal{F}\mathcal{B}}A^*_tA_t'+
\frac{A^*_{x}A_{x}'+A^*_{y}A_{y}'}{1+B_z^2u^4}+\frac{1}{\mathcal{H}}A^*_zA_z'\right),\\
S_{\epsilon}&=&-2K_{D7}\int \frac{d^4k}{(2\pi)^4}\frac{Q_{B_y}}{u}\left(-\frac{1}{\mathcal{F}\mathcal{B}}A^*_tA_t'+
A^*_{x}A_{x}'+\frac{A^*_yA_y'+A^*_zA_z'}{\mathcal{H}+B_y^2u^4}\right),
\end{eqnarray}  
where
\begin{eqnarray}
Q_{B_z}&=&\nonumber\frac{(1-\psi^2)^2\mathcal{F}\sqrt{\mathcal{B}(1+B_z^2u^4)}e^{-\frac{3\phi}{4}}}
{\sqrt{1-\psi^2+u^2\mathcal{F}e^{\frac{1}{2}\phi}\psi^{\prime 2}}},\\
Q_{B_y}&=&\frac{(1-\psi^2)^2\mathcal{F}\sqrt{\mathcal{B}(\mathcal{H}+B_z^2u^4)}e^{-\frac{\phi}{4}}}
{\sqrt{1-\psi^2+u^2\mathcal{F}e^{\frac{1}{2}\phi}\psi^{\prime 2}}}.
\end{eqnarray}
The field equations of the gauge fields here also take the Maxwell form,   
\begin{eqnarray}\label{Maxwellform}
\partial_{\mu}(\sqrt{-\mbox{det}G_{\mu\nu}}e^{-\phi} G^{\mu\alpha}G^{\nu\beta}F_{\alpha\beta})=0.
\end{eqnarray}
The equations above can be converted into gauge-invariant forms from the general expressions in Appendix A. Here we list the diagonal terms of the induced metric pertinent to the computations,
\begin{eqnarray}\nonumber
G^{tt}&=&-\frac{u^2}{\mathcal{F}\mathcal{B}},\quad G^{xx}=G^{yy}=\frac{u^2}{1+B_z^2u^4},\\
G^{zz}&=&\frac{u^2}{\mathcal{H}}, \quad G^{uu}=\frac{u^2\mathcal{F}(1-\psi^2)}{1-\psi^2+u^2\mathcal{F}e^{\frac{\phi}{2}}\psi'^2}
\end{eqnarray}
for $B_z\neq 0$ and 
\begin{eqnarray}\nonumber
G^{tt}&=&-\frac{u^2}{\mathcal{F}\mathcal{B}},\quad G^{xx}=\mathcal{H}G^{zz}=\frac{u^2\mathcal{H}}{\mathcal{H}+B_y^2u^4},\\
G^{yy}&=&u^2, \quad G^{uu}=\frac{u^2\mathcal{F}(1-\psi^2)}{1-\psi^2+u^2\mathcal{F}e^{\frac{\phi}{2}}\psi'^2}
\end{eqnarray}
for $B_y\neq 0$.
After solving the field equations, we can follow the same procedure as introduced in the previous section to compute the spectral functions of photons. To compare the results in the presence of magnetic field and anisotropy, we have to fix the temperature and quark mass in different setups. We firstly consider the situation when the magnetic field points along the anisotropic direction for which the results are shown in Fig.\ref{isoBzanisoBz}, Fig.\ref{isoBzanisoBzqxAy}, and Fig.\ref{isoBzanisoBzqxAz}, where the spectral functions are in the unit of $\pi T$. As shown in Fig.\ref{isoBzanisoBz}, the spectra for photons emitted parallel to the magnetic field are suppressed at small $\hat{\omega}$, while they saturate the isotropic spectra in the absence of magnetic field at large $\hat{\omega}$. When further incorporating the pressure anisotropy, the spectra for photons emitted parallel to the anisotropic direction are enhanced, which is similar to the scenario in the absence of magnetic field as shown in the previous section; their amplitudes surpass the isotropic ones with zero magnetic field at large $\hat{\omega}$. 

For the photons emitted perpendicular to the magnetic field, as shown in Fig.\ref{isoBzanisoBzqxAy} and Fig.\ref{isoBzanisoBzqxAz}, their spectra are enhanced at large $\hat{\omega}$. Also, the anisotropic effect make no drastic modifications to the spectra at large $\hat{\omega}$. However, at moderate $\hat{\omega}$, a resonance emerges in the spectrum led by heavy quarks for photons moving perpendicular to the magnetic field. The resonance is more prominent when the photos are polarized parallel to the magnetic field as illustrated by the dashed blue curve in Fig.\ref{isoBzanisoBzqxAz}. When further incorporating the pressure anisotropy, the resonance is smoothed out. 

\begin{figure}[h]
\begin{minipage}{7cm}
\begin{center}
{\includegraphics[width=7.5cm,height=5cm,clip]{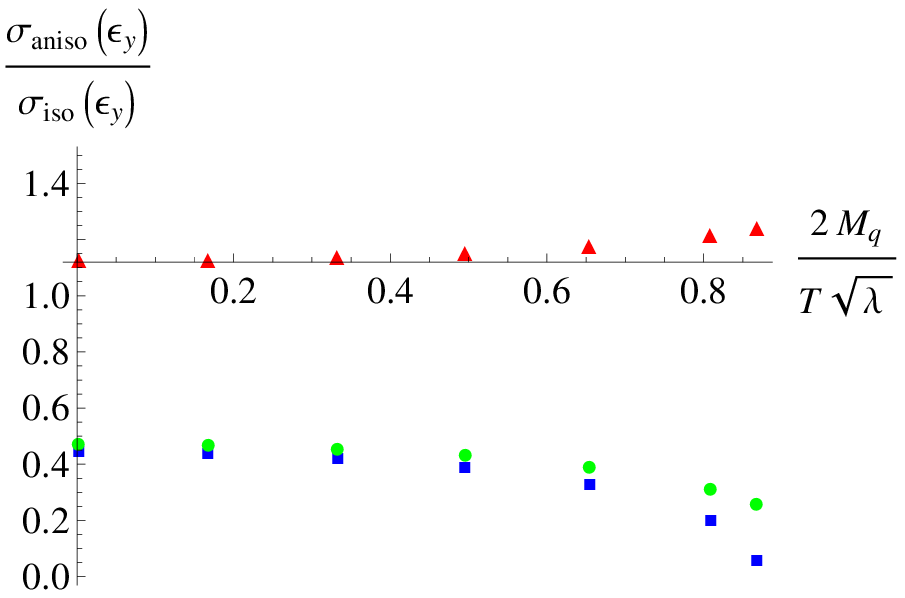}}
\caption{The ratios of DC conductivity with $\epsilon_T=\epsilon_y$ versus quark mass. The red(triangle), green(circle), and blue(square) dots correspond to $(a/T,B_z/(\pi T)^2)=(4.4,0)$, $(0,2)$, and $(4.4,2)$, respectively.}\label{ecAy}
\end{center}
\end{minipage}
\hspace {1cm}
\begin{minipage}{7cm}
\begin{center}
{\includegraphics[width=7.5cm,height=5cm,clip]{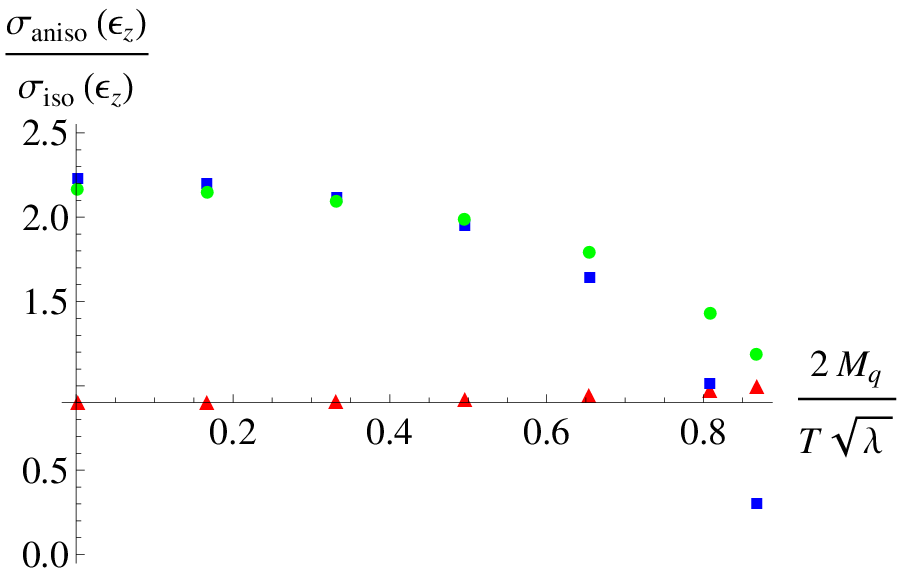}}
\caption{The ratios of DC conductivity with $\epsilon_T=\epsilon_z$ versus quark mass. The red(triangle), green(circle), and blue(square) dots correspond to $(a/T,B_z/(\pi T)^2)=(4.4,0)$, $(0,2)$, and $(4.4,2)$, respectively.}
\label{ecAz}
\end{center}
\end{minipage}
\end{figure}

In the zero-frequency limit, we can also evaluate the DC conductivity by employing (\ref{Dcconductivity}), where the results are shown in Fig.\ref{ecAy} and Fig.\ref{ecAz}. As illustrated in Fig.\ref{ecAy}, compared to the isotropic case in the absence of magnetic field, we find that the conductivity for photons with the polarization perpendicular to the anisotropic direction is enhanced in particular for the embedding with heavy quarks. On the contrary, the conductivity for the polarization perpendicular to the magnetic field is suppressed. When the quark mass is increased, the suppression becomes more robust. In contrast, as illustrated in Fig.\ref{ecAz}, for photons with the polarization along the anisotropic direction, the conductivity is almost unchanged compared to the isotropic one except for the embedding with heavy quarks. However, the conductivity for the photons with the polarization parallel to the magnetic field is larger than the isotropic one for the embedding with light quarks. When the quark mass is increased, the enhancement monotonically decreases and even turns into suppression when approaching the critical mass.          

\begin{figure}[h]
\begin{minipage}{7cm}
\begin{center}
{\includegraphics[width=7.5cm,height=5cm,clip]{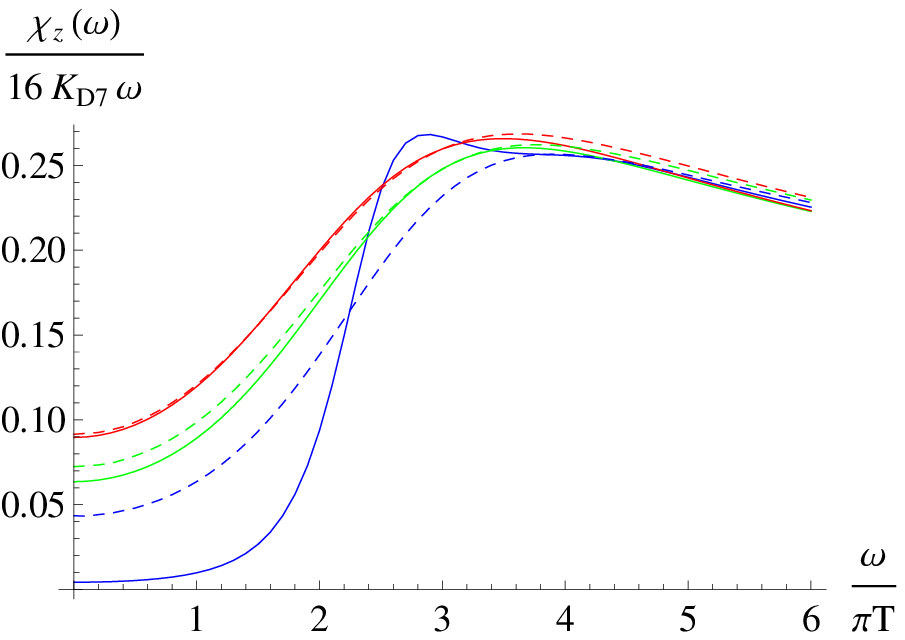}}
\caption{The spectral functions with $k=(-\omega,\omega,0,0)$ and $\epsilon_T=\epsilon_z$.
The solid and dashed curves correspond to $(a/T, B_y/(\pi T)^2)=(0,2)$ and $(4.4,2)$, respectively.}\label{chianisoByqxAz}
\end{center}
\end{minipage}
\hspace {1cm}
\begin{minipage}{7cm}
\begin{center}
{\includegraphics[width=7.5cm,height=5cm,clip]{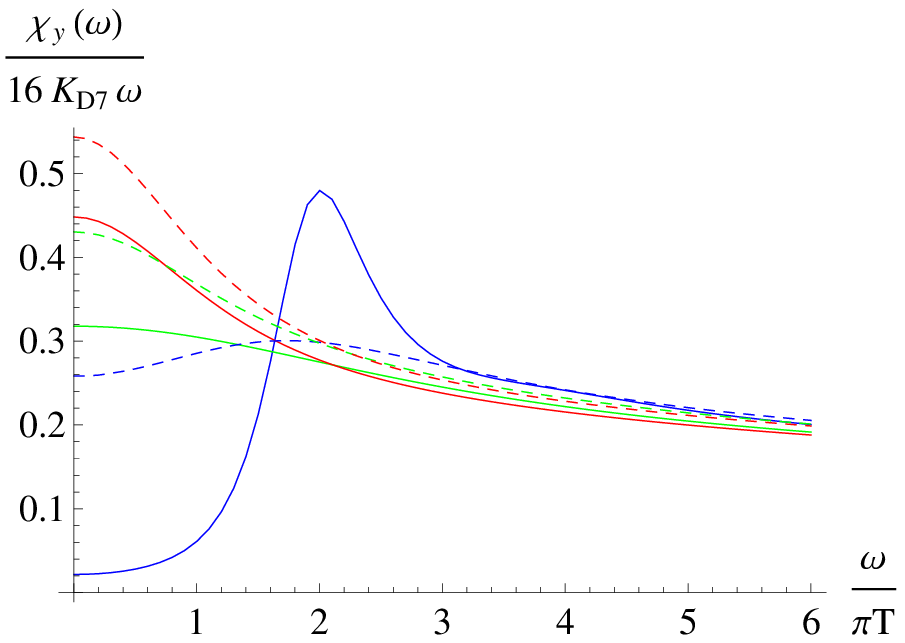}}
\caption{The spectral functions with $k=(-\omega,\omega,0,0)$ and $\epsilon_T=\epsilon_y$.
The solid and dashed curves correspond to $(a/T, B_y/(\pi T)^2)=(0,2)$ and $(4.4,2)$, respectively.}
\label{chianisoByqxAy}
\end{center}
\end{minipage}
\end{figure}

\begin{figure}[h]
\begin{minipage}{7cm}
\begin{center}
{\includegraphics[width=7.5cm,height=5cm,clip]{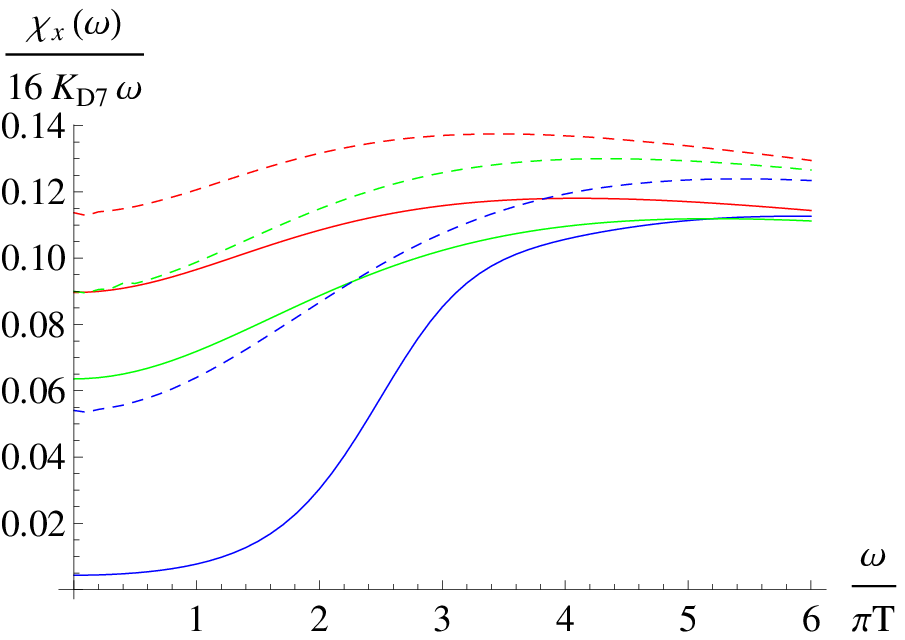}}
\caption{The spectral functions with $k=(-\omega,0,\omega,0)$ and $\epsilon_T=\epsilon_x$.
The solid and dashed curves correspond to $(a/T, B_y/(\pi T)^2)=(0,2)$ and $(4.4,2)$, respectively.}\label{chianisoByqyAx}
\end{center}
\end{minipage}
\hspace {1cm}
\begin{minipage}{7cm}
\begin{center}
{\includegraphics[width=7.5cm,height=5cm,clip]{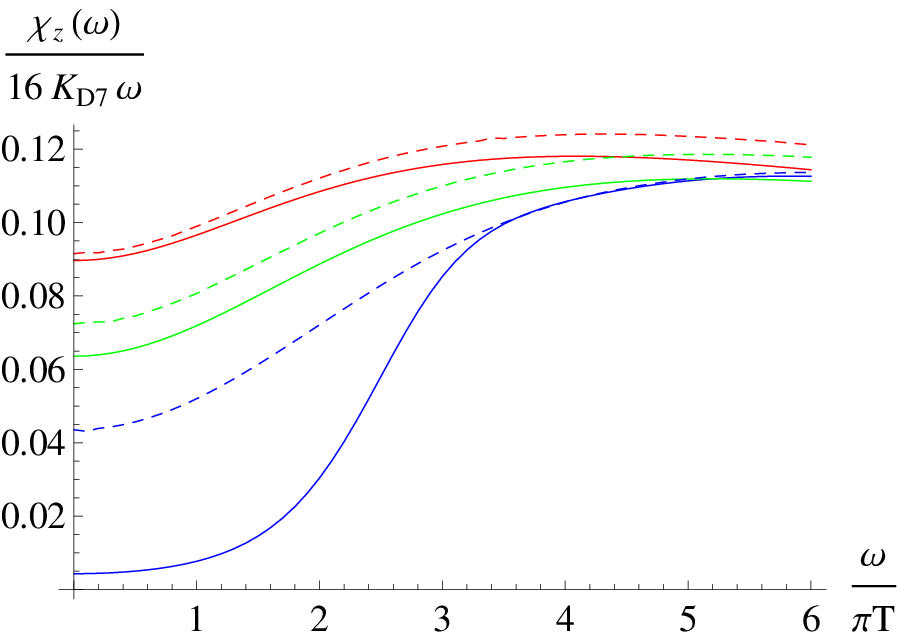}}
\caption{The spectral functions with $k=(-\omega,0,\omega,0)$ and $\epsilon_T=\epsilon_z$.
The solid and dashed curves correspond to $(a/T, B_y/(\pi T)^2)=(0,2)$ and $(4.4,2)$, respectively.}
\label{chianisoByqyAz}
\end{center}
\end{minipage}
\end{figure}

When the magnetic field and anisotropy coexist and point perpendicular to each other, the rotational symmetry is fully broken. The photons moving in distinct directions with different polarizations will lead to a variety of spectra, but the general features are not particularly altered from what we have discussed in the paragraph above. The spectra for photons moving parallel to the anisotropic direction and perpendicular to the magnetic field receive the maximum enhancement at large frequency. When the photons are emitted perpendicular to the magnetic field in the isotropic medium, the resonance appears for the spectra with heavy quarks at moderate frequency. The presence of anisotropy then smooths out the resonance regardless of the moving directions of photons.     
We merely present the results in 
Fig.\ref{chianisoByqxAz}-\ref{chianisoByqyAz} for references, where the correspondences between the colors of curves and different values of quark mass are the same as those in Fig.\ref{isoBzanisoBz}.

\begin{figure}[h]
\begin{center}
{\includegraphics[width=7.5cm,height=5cm,clip]{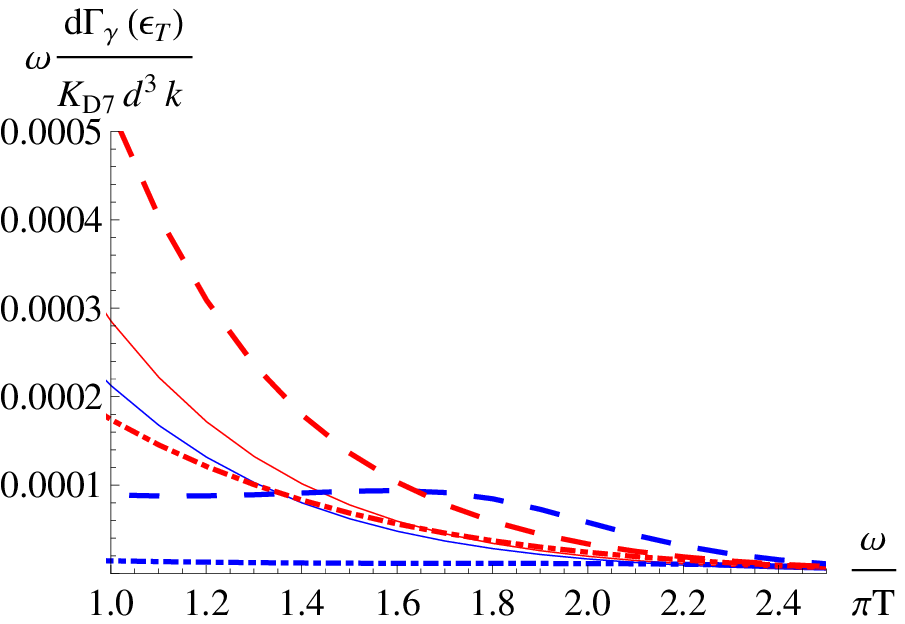}}
\caption{The red and blue (upper and lower at $\omega/(\pi T)=1$) curves represent the differential emission rate per unit volume with $k=(-\omega,\omega,0,0)$ for $\hat{M}_q/(\pi T)=0.45$ and $0.86$. The solid, dashed, and dot-dashed ones correspond to $(\epsilon_T,B_z/(\pi T)^2)=(\epsilon_{z(y)},0)$, $(\epsilon_z,2)$, and $(\epsilon_y,2)$, respectively.}\label{resonance}
\end{center}
\end{figure}

\section{\label{sec:level1}Discussion}
In this paper, we have evaluated the thermal photon spectra originated from massive quarks in the anisotropic plasma with moderate pressure anisotropy and constant yet strong magnetic field through the holographic approach.
At large frequency, we found that the amplitudes of spectra with different quark mass are increased by the magnetic field when the photons move perpendicular to it. The spectra for photons moving parallel to the magnetic field saturate the results in the absence of magnetic field. In the presence of pressure anisotropy, the spectra for photons moving along the anisotropy direction are enhanced, while the ones moving perpendicular to the anisotropic direction receive no enhancement compared to the isotropic results at large frequency. At moderate frequency, the magnetic field triggers a resonance for the spectrum with heavy quarks when the photons move perpendicular to the magnetic field. The resonance becomes more robust when the photons are polarized along the magnetic field. However, the pressure anisotropy leads to an competing effect and suppresses the resonance. 

\begin{figure}[h]
\begin{minipage}{7cm}
\begin{center}
{\includegraphics[width=7.5cm,height=5cm,clip]{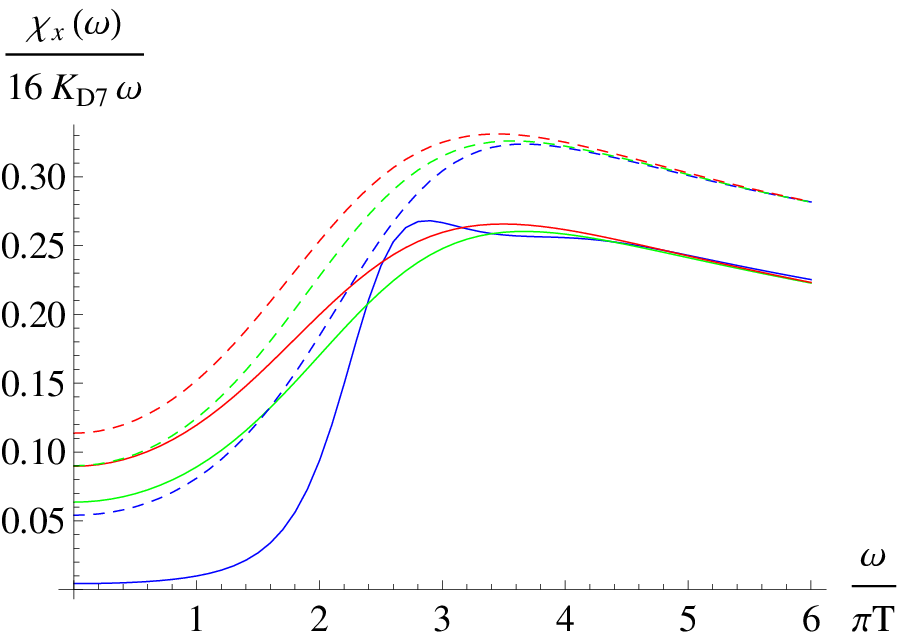}}
\caption{The spectral functions with $k=(-\omega,0,0,\omega)$ and $\epsilon_T=\epsilon_x$.
The solid and dashed curves correspond to $(a/T, B_y/(\pi T)^2)=(0,2)$ and $(4.4,2)$, respectively.}\label{chianisoByqzAx}
\end{center}
\end{minipage}
\hspace {1cm}
\begin{minipage}{7cm}
\begin{center}
{\includegraphics[width=7.5cm,height=5cm,clip]{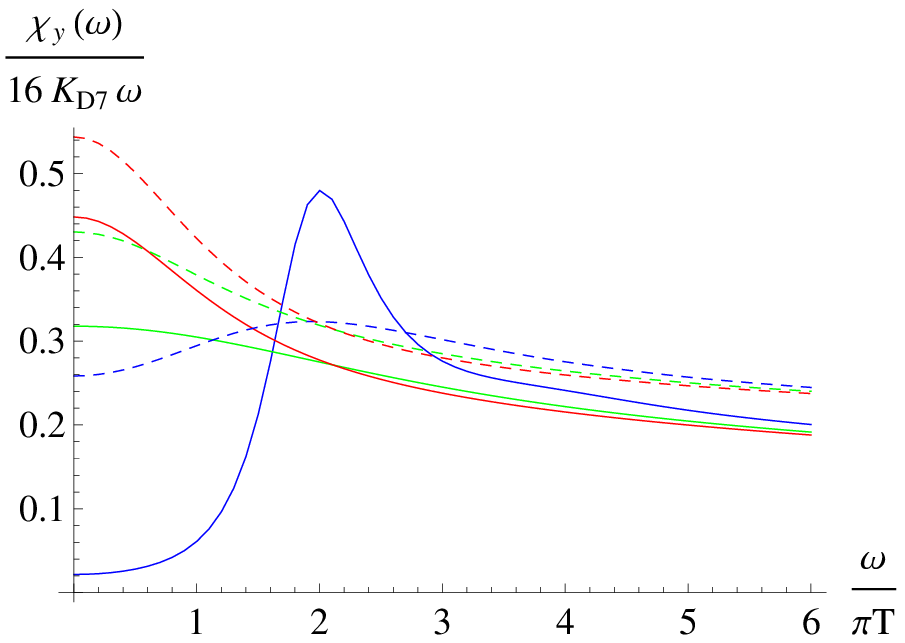}}
\caption{The spectral functions with $k=(-\omega,0,0,\omega)$ and $\epsilon_T=\epsilon_y$.
The solid and dashed curves correspond to $(a/T, B_y/(\pi T)^2)=(0,2)$ and $(4.4,2)$, respectively.}
\label{chianisoByqzAy}
\end{center}
\end{minipage}
\end{figure}

The enhancements of the photon spectra with massive quarks at large frequency led by anisotropy and magnetic field are somewhat expected since the enhancements have been found in the limit of massless quarks \cite{Patino:2012py,PhysRevD.87.026005}. When the energy of photons dominates the quark mass, the contributions from massive and massless quarks should degenerate. The enhancements will persist for $\hat{\omega}\rightarrow\infty$. In the coexistence of both effects, the enhancement of the spectra could be further amplified. As illustrated in Fig.\ref{chianisoByqzAx} and Fig.\ref{chianisoByqzAy}, for which the magnetic field is perpendicular to the anisotropy direction, the spectra at large frequency in the presence of magnetic field can be further enhanced by anisotropy. When the photons move along the anisotropic direction in the above situation, their amplitudes of the spectra can be maximally enhanced. Such a scenario may be in analogy to the photon production in relativistic heavy ion collisions. In the peripheral collisions in experiments, the orientation of averaged magnetic field generated by two colliding nuclei should be perpendicular to the reaction plane. The initial geometry of the medium will lead to pressure anisotropy, in which the largest pressure is perpendicular to the averaged magnetic field and to the beam direction.
At mid rapidity, the observation of large elliptic flow of direct photons corresponds to the excessive production of photons along the orientation with largest pressure. In our model, the maximum enhancement of the thermal photons produced along the anisotropic direction and perpendicular to the magnetic field may qualitatively suggest the cause of such large flow of direct photons.

Nonetheless, there exist caveats when making the comparison above beyond the difference in the SYM theory and QCD. Firstly, the pressure anisotropy in the MT model is distinct from the one in QGP in directions. In the MT model, the rotational symmetry is still preserved on the plane perpendicular to the anisotropic direction with larger pressure, which is drastically distinct from the QGP that the rotational symmetry should be fully broken. Second, the medium will gradually expand and thus the temperature of QGP could be spacetime dependent. Furthermore, the magnetic field generated by the colliding nuclei should decay rapidly with time. The lifetime modified by the presence of matters could be controversial, where the different estimations can be found in \cite{Tuchin:2013ie,Tuchin:2013apa} and \cite{McLerran:2013hla}. Despite the distinction between our model and QGP in reality, the results from such a toy model still provides us qualitative understandings of the influence of pressure anisotropy and magnetic filed on the photon production in the strongly coupled scenario.

At moderate frequency, the spectra with heavy quarks in the presence of magnetic field and anisotropy are rather intriguing. As indicated in \cite{CasalderreySolana:2008ne}, the resonance appearing in photon spectra may implies the decay of heavy mesons to on-shell photons. In the previous study of photon spectra with massive quarks in the absence of magnetic field in an isotropic medium \cite{Mateos:2007yp}, it has been shown that the resonance starts to emerge when the quark mass approaches the critical mass. Since the presence of magnetic field reduces the critical mass, the resonance appears for the spectrum with smaller quark mass, which suggests that the decay from lighter mesons to on-shell photons will be accessible with the aid of magnetic field.
Moreover, the dependence on orientation and polarization with respect to the magnetic field makes the resonance distinguishable from the isotropic one triggered by the increase of quark mass in the absence of magnetic field \cite{Mateos:2007yp}.

Here we plot the differential emission rate per unit volume in the unit of $(\pi T)^2$ in Fig.\ref{resonance} around moderate frequency, where the resonance from heavy quarks could be comparable to the spectrum with lighter quarks in this regime. In the presence of magnetic field, the impact of the resonance on the shape of the spectrum at moderate frequency is even more pronounced than the enhancement at large frequency, which could generate a mild peak of $v_2$ at moderate frequency.     
Despite the over-simplification of our model, the orientation-dependence resonance could give a rise to the mild peak of $v_2$ in the intermediate energy. On the other hand, it is also indicated in \cite{Yee:2013qma} that the photons with out-plane polarizations, which correspond to the photons polarized along the magnetic field when moving perpendicular to the field in our setup, will account for the primary contributions beyond small frequency. The polarization dependence of the enhanced spectra of direct photons could be substantial to clarify the cause of large $v_2$ in future experiments.

On the contrary, the further inclusion of pressure anisotropy increases the critical mass and thus reduces the resonance in our model, which favors the meson melting in the plasma. Our findings is consistent with \cite{Giataganas:2012zy,Chernicoff:2012bu,Chakraborty:2012dt}, in which the increase of anisotropy results in the decrease of the screening length of the quark-antiquark potential in the MT geometry at fixed temperature. At first glance, it is surprising that the qualitative features of heavy-meson suppression led by pressure anisotropy found in the MT geometry are contradictory to those obtained in weakly coupled approaches \cite{Dumitru:2007hy,Dumitru:2009fy} and anisotropic hydrodynamics \cite{Strickland:2011aa,Strickland:2011mw,Mocsy:2013syh} for anisotropic QCD plasmas, 
where the anisotropic media result in less suppression for heavy mesons in comparison with the isotropic ones. Nonetheless, one should recall the difference between the setup in the MT model and in those approaches more analogous to QGP in reality. In the MT model, the ratio of shear viscosity to entropy density is decreased by the increase of pressure anisotropy \cite{Rebhan:2011vd}. This is drastically distinct from the approaches related to QGP, in which the pressure anisotropy makes the plasma more viscous.    
As a result, the anisotropic effect in the MT model facilitates the deconfinement while the anisotropy in the viscous QGP favors the confinement of heavy mesons. We thus conclude that the resonance stemming from the presence of magnetic field may not be suppressed by pressure anisotropy in QGP.

Even though the MT model with constant field may not match some qualitative properties of the QGP in reality, the model is of considerable interest in its own right. To acquire better understandings of the competing effect between the anisotropy and magnetic field on the meson melting in the strongly coupled SYM plasma with fundamental matters, it is rather intriguing to further investigate the meson spectral functions and thermodynamics in Minkowski embeddings \cite{wuyang}. From phenomenological perspectives, it is as well crucial to incorporate time-dependent magnetic field in future works. In addition, to manifest the influence of the resonance originated from the heavy quarks on photon production, the computation of $v_2$ for the thermal photons with magnetic field in the D3/D7 system will be reported in \cite{wuyangv2}. Finally, we expect the embedding of probe flavor brane is stable, but it needs further considerations.

\section{\label{sec:level1}Acknowledgement}

The authors thank E. Caceres, A. Kundu, and B. M\"uller for useful discussions. This material is based upon work supported
by DOE grants DE-FG02-05ER41367, de-sc0005396 (D.~L.~Yang), the National Science Council (NSC
101-2811-M-009-015) and the Nation Center for Theoretical Science, Taiwan (S.~Y.~Wu).

\section{\label{sec:level1}Appendix}
\subsection{\label{sec:level1}General Expressions for Field Equations}
In this appendix, we demonstrate the derivation of general expressions of field equations in the string frame in gauge invariant forms. From the quadratic term of the field strength in the DBI action, we have the field equations in the Maxwell form as shown in (\ref{Maxwellform}). By taking Fourier transform of the gauge field as shown in (\ref{FTforA}), the field equations in the gauge of $A_u=0$ now read
\begin{eqnarray}\label{Maxgenral}
\nonumber&&(MG^{uu}G^{jj}A_j')'-MG^{jj}(G^{tt}\omega^2+G^{ii}q^2)A_j=0,\\
\nonumber&&(MG^{uu}G^{tt}A_t')'-MG^{tt}G^{ii}(q^2A_t+q\omega A_i)=0,\\
\nonumber&&(MG^{uu}G^{tt}A_i')'-MG^{tt}G^{ii}(\omega^2A_i+q\omega A_t)=0,\\
&&\omega G^{tt}A_t'-qG^{ii}A_i'=0,
\end{eqnarray}    
where $M=e^{-\phi}\sqrt{-\mbox{det}G_{\mu\nu}}$ and $i$ denotes the propagating direction and $j$ denotes the polarization of the gauge field. The first equation here represents the transverse mode and the rest three contribute to the longitudinal modes. To rewrite the field equations into the gauge invariant form, we define 
\begin{eqnarray}\label{Efield}
E_i=qA_t+\omega A_i,\quad E_j=\omega A_j,
\end{eqnarray}   
where we set $k_0=-\omega$ and $k_i=q$. By using the first equation of (\ref{Maxgenral}) and (\ref{Efield}), we obtain the gauge invariant form for the transverse modes,
\begin{eqnarray}
E_j''+(\log(MG^{uu}G^{jj}))'E_j'-\frac{1}{G^{uu}}(G^{tt}\omega^2+G^{ii}q^2)E_j=0.
\end{eqnarray} 
By combining the rest three equations in (\ref{Maxgenral}) and (\ref{Efield}) and doing some algebras, we then derive the gauge invariant form for the longitudinal modes,
\begin{eqnarray}
E_i''+\left[(\log(MG^{uu}G^{ii}))'+\left(\log\left(\frac{G^{tt}}{G^{ii}}\right)\right)'\frac{G^{ii}q^2}{k^2}\right]E_i'
-\frac{k^2}{G^{uu}}E_i=0,
\end{eqnarray}
where $k^2=G^{tt}\omega^2+G^{ii}q^2$.  
Furthermore, by using the last equation in (\ref{Maxgenral}), we can also rewrite the near-boundary action into the gauge invariant form as
\begin{eqnarray}\nonumber
S_{\epsilon}&=&-2K_{D7}\int \frac{d^4k}{(2\pi)^4}G^{uu}M\left(G^{tt}A^*_tA_t'+G^{jj}A^*_jA_j'+G^{ii}A^*_iA_i'\right)\\
&=&-2K_{D7}\int \frac{d^4k}{(2\pi)^4}G^{uu}M\left(\frac{G^{ii}G^{tt}}{q^2G^{ii}+\omega^2G^{tt}}E_i^*E_i'+\frac{G^{jj}}{\omega^2}E_j^*E_j'\right).
\end{eqnarray} 
Finally, by implementing (\ref{chiC}), we obtain the photon spectral density
\begin{eqnarray}\label{gchi}
\chi_{\epsilon_j}(\omega)=8K_{D7}\mbox{Im}\lim_{u\rightarrow 0}\frac{G^{uu}MG^{jj}E'_j(u,\omega)}{E_j(u,\omega)}.
\end{eqnarray}   

\subsection{\label{sec:level1}Near-Boundary Expansion}
To analyze the asymptotic behavior of $\psi(u)$ with black-hole embedding near the boundary in MT metric, we may consider the situation with small anisotropy since the leading-order expansion of the MT geometry in terms of $a/T$ can be solved analytically. In this limit for $a/T<\ll 1$, the leading-order solution of MT geometry reads \cite{Mateos:2011tv},
\begin{eqnarray}\label{comp}
\nonumber\mathcal{F}(u)&=&1-\frac{u^4}{u_h^4}+a^2\mathcal{F}_2(u)+\mathcal{O}(a^4),\\
\nonumber\mathcal{B}(u)&=&1+a^2\mathcal{B}_2(u)+\mathcal{O}(a^4),\\
\mathcal{H}(u)&=&e^{-\phi(u)},\mbox{}\phi(u)=a^2\phi_2(u)+\mathcal{O}(a^4),
\end{eqnarray}
where the coefficients for the anisotropic contributions are given by
\begin{eqnarray}
\nonumber\mathcal{F}_2(u)&=&\frac{1}{24u_h^2}\left[8u^2(u_h^2-u^2)-10u^4\mbox{log}2+(3u_h^4+7u^4)\mbox{log}\left(1+\frac{u^2}{u_h^2}\right)\right],\\
\nonumber\mathcal{B}_2(u)&=&-\frac{u_h^2}{24}\left[\frac{10u^2}{u_h^2+u^2}+\mbox{log}\left(1+\frac{u^2}{u_h^2}\right)\right],\\
\phi_2(u)&=&-\frac{u_h^2}{4}\mbox{log}\left(1+\frac{u^2}{u_h^2}\right).
\end{eqnarray}
We then insert the analytic expression of the background metric above into (\ref{psiEOM}) and solve for $\psi(u)$ near the boundary. We may assume that the asymptotic expansion of $\psi(u)$ takes form,
\begin{eqnarray}
\psi(u)=\psi_1\frac{u}{u_h}+\psi_3\frac{u^3}{u_h^3}+\psi_5\frac{u^5}{u_h^5}+
a^2\log\left(\frac{u}{u_h}\right)\left(\rho_3\frac{u^3}{u_h^3}+\rho_5\frac{u^5}{u_h^5}\right)+\dots,
\end{eqnarray}  
where the logarithmic terms come from anisotropy and have to vanish as $a\rightarrow 0$. 
Up to the $\mathcal{O}(a^2)$ and $\mathcal{O}(u^6)$ of (\ref{psiEOM}), we find
\begin{eqnarray}\nonumber
\psi_5&=&\frac{1}{8}\psi_1(1 + 8\psi_1 \psi_3)+\frac{a^2}{96}(8\psi_1^3 -9\psi_3+ \psi_1(-2 + \log 32)),\\
\rho_3&=&\frac{5}{24}\psi_1,\quad\rho_5=\psi_1^2\rho_3,
\end{eqnarray} 
where the coefficients of higher-order terms are determined by $\psi_1$ and $\psi_3$ associated with the quark mass and condensate, respectively. Since the leading-order logarithmic term dominates the $\mathcal{O}(u^3)$ term near the boundary, it is awkward to extract $\psi_3$ in terms of the expansion in the $u$ coordinate when the background geometry is anisotropic.      

\begin{thebibliography}{65}
\expandafter\ifx\csname natexlab\endcsname\relax\def\natexlab#1{#1}\fi
\expandafter\ifx\csname bibnamefont\endcsname\relax
  \def\bibnamefont#1{#1}\fi
\expandafter\ifx\csname bibfnamefont\endcsname\relax
  \def\bibfnamefont#1{#1}\fi
\expandafter\ifx\csname citenamefont\endcsname\relax
  \def\citenamefont#1{#1}\fi
\expandafter\ifx\csname url\endcsname\relax
  \def\url#1{\texttt{#1}}\fi
\expandafter\ifx\csname urlprefix\endcsname\relax\def\urlprefix{URL }\fi
\providecommand{\bibinfo}[2]{#2}
\providecommand{\eprint}[2][]{\url{#2}}

\bibitem[{\citenamefont{Adare et~al.}(2010{\natexlab{a}})}]{Adare:2008ab}
\bibinfo{author}{\bibfnamefont{A.}~\bibnamefont{Adare}} \bibnamefont{et~al.}
  (\bibinfo{collaboration}{PHENIX Collaboration}),
  \bibinfo{journal}{Phys.Rev.Lett.} \textbf{\bibinfo{volume}{104}},
  \bibinfo{pages}{132301} (\bibinfo{year}{2010}{\natexlab{a}}),
  \eprint{0804.4168}.

\bibitem[{\citenamefont{Adare et~al.}(2010{\natexlab{b}})}]{Adare:2009qk}
\bibinfo{author}{\bibfnamefont{A.}~\bibnamefont{Adare}} \bibnamefont{et~al.}
  (\bibinfo{collaboration}{PHENIX Collaboration}), \bibinfo{journal}{Phys.Rev.}
  \textbf{\bibinfo{volume}{C81}}, \bibinfo{pages}{034911}
  (\bibinfo{year}{2010}{\natexlab{b}}), \eprint{0912.0244}.

\bibitem[{\citenamefont{Gale}(2009)}]{Gale:2009gc}
\bibinfo{author}{\bibfnamefont{C.}~\bibnamefont{Gale}} (\bibinfo{year}{2009}),
  \eprint{0904.2184}.

\bibitem[{\citenamefont{Adare et~al.}(2012)}]{Adare:2011zr}
\bibinfo{author}{\bibfnamefont{A.}~\bibnamefont{Adare}} \bibnamefont{et~al.}
  (\bibinfo{collaboration}{PHENIX Collaboration}),
  \bibinfo{journal}{Phys.Rev.Lett.} \textbf{\bibinfo{volume}{109}},
  \bibinfo{pages}{122302} (\bibinfo{year}{2012}), \eprint{1105.4126}.

\bibitem[{\citenamefont{Lohner}(2012)}]{Lohner:2012ct}
\bibinfo{author}{\bibfnamefont{D.}~\bibnamefont{Lohner}}
  (\bibinfo{collaboration}{ALICE Collaboration}) (\bibinfo{year}{2012}),
  \eprint{1212.3995}.

\bibitem[{\citenamefont{Kapusta et~al.}(1991)\citenamefont{Kapusta, Lichard,
  and Seibert}}]{Kapusta:1991qp}
\bibinfo{author}{\bibfnamefont{J.~I.} \bibnamefont{Kapusta}},
  \bibinfo{author}{\bibfnamefont{P.}~\bibnamefont{Lichard}}, \bibnamefont{and}
  \bibinfo{author}{\bibfnamefont{D.}~\bibnamefont{Seibert}},
  \bibinfo{journal}{Phys.Rev.} \textbf{\bibinfo{volume}{D44}},
  \bibinfo{pages}{2774} (\bibinfo{year}{1991}).

\bibitem[{\citenamefont{Baier et~al.}(1992)\citenamefont{Baier, Nakkagawa,
  Niegawa, and Redlich}}]{Baier:1991em}
\bibinfo{author}{\bibfnamefont{R.}~\bibnamefont{Baier}},
  \bibinfo{author}{\bibfnamefont{H.}~\bibnamefont{Nakkagawa}},
  \bibinfo{author}{\bibfnamefont{A.}~\bibnamefont{Niegawa}}, \bibnamefont{and}
  \bibinfo{author}{\bibfnamefont{K.}~\bibnamefont{Redlich}},
  \bibinfo{journal}{Z.Phys.} \textbf{\bibinfo{volume}{C53}},
  \bibinfo{pages}{433} (\bibinfo{year}{1992}).

\bibitem[{\citenamefont{Aurenche et~al.}(1998)\citenamefont{Aurenche, Gelis,
  Kobes, and Zaraket}}]{Aurenche:1998nw}
\bibinfo{author}{\bibfnamefont{P.}~\bibnamefont{Aurenche}},
  \bibinfo{author}{\bibfnamefont{F.}~\bibnamefont{Gelis}},
  \bibinfo{author}{\bibfnamefont{R.}~\bibnamefont{Kobes}}, \bibnamefont{and}
  \bibinfo{author}{\bibfnamefont{H.}~\bibnamefont{Zaraket}},
  \bibinfo{journal}{Phys.Rev.} \textbf{\bibinfo{volume}{D58}},
  \bibinfo{pages}{085003} (\bibinfo{year}{1998}), \eprint{hep-ph/9804224}.

\bibitem[{\citenamefont{Arnold et~al.}(2001)\citenamefont{Arnold, Moore, and
  Yaffe}}]{Arnold:2001ms}
\bibinfo{author}{\bibfnamefont{P.~B.} \bibnamefont{Arnold}},
  \bibinfo{author}{\bibfnamefont{G.~D.} \bibnamefont{Moore}}, \bibnamefont{and}
  \bibinfo{author}{\bibfnamefont{L.~G.} \bibnamefont{Yaffe}},
  \bibinfo{journal}{JHEP} \textbf{\bibinfo{volume}{0112}}, \bibinfo{pages}{009}
  (\bibinfo{year}{2001}), \eprint{hep-ph/0111107}.

\bibitem[{\citenamefont{Arnold et~al.}(2002)\citenamefont{Arnold, Moore, and
  Yaffe}}]{Arnold:2002ja}
\bibinfo{author}{\bibfnamefont{P.~B.} \bibnamefont{Arnold}},
  \bibinfo{author}{\bibfnamefont{G.~D.} \bibnamefont{Moore}}, \bibnamefont{and}
  \bibinfo{author}{\bibfnamefont{L.~G.} \bibnamefont{Yaffe}},
  \bibinfo{journal}{JHEP} \textbf{\bibinfo{volume}{0206}}, \bibinfo{pages}{030}
  (\bibinfo{year}{2002}), \eprint{hep-ph/0204343}.

\bibitem[{\citenamefont{Ghiglieri et~al.}(2013)\citenamefont{Ghiglieri, Hong,
  Kurkela, Lu, Moore et~al.}}]{Ghiglieri:2013gia}
\bibinfo{author}{\bibfnamefont{J.}~\bibnamefont{Ghiglieri}},
  \bibinfo{author}{\bibfnamefont{J.}~\bibnamefont{Hong}},
  \bibinfo{author}{\bibfnamefont{A.}~\bibnamefont{Kurkela}},
  \bibinfo{author}{\bibfnamefont{E.}~\bibnamefont{Lu}},
  \bibinfo{author}{\bibfnamefont{G.~D.} \bibnamefont{Moore}},
  \bibnamefont{et~al.} (\bibinfo{year}{2013}), \eprint{1302.5970}.

\bibitem[{\citenamefont{Maldacena}(1998)}]{Maldacena:1997re}
\bibinfo{author}{\bibfnamefont{J.~M.} \bibnamefont{Maldacena}},
  \bibinfo{journal}{Adv.Theor.Math.Phys.} \textbf{\bibinfo{volume}{2}},
  \bibinfo{pages}{231} (\bibinfo{year}{1998}), \eprint{hep-th/9711200}.

\bibitem[{\citenamefont{Witten}(1998{\natexlab{a}})}]{Witten:1998qj}
\bibinfo{author}{\bibfnamefont{E.}~\bibnamefont{Witten}},
  \bibinfo{journal}{Adv.Theor.Math.Phys.} \textbf{\bibinfo{volume}{2}},
  \bibinfo{pages}{253} (\bibinfo{year}{1998}{\natexlab{a}}),
  \eprint{hep-th/9802150}.

\bibitem[{\citenamefont{Gubser et~al.}(1998)\citenamefont{Gubser, Klebanov, and
  Polyakov}}]{Gubser:1998bc}
\bibinfo{author}{\bibfnamefont{S.}~\bibnamefont{Gubser}},
  \bibinfo{author}{\bibfnamefont{I.~R.} \bibnamefont{Klebanov}},
  \bibnamefont{and} \bibinfo{author}{\bibfnamefont{A.~M.}
  \bibnamefont{Polyakov}}, \bibinfo{journal}{Phys.Lett.}
  \textbf{\bibinfo{volume}{B428}}, \bibinfo{pages}{105} (\bibinfo{year}{1998}),
  \eprint{hep-th/9802109}.

\bibitem[{\citenamefont{Aharony et~al.}(2000)\citenamefont{Aharony, Gubser,
  Maldacena, Ooguri, and Oz}}]{Aharony:1999ti}
\bibinfo{author}{\bibfnamefont{O.}~\bibnamefont{Aharony}},
  \bibinfo{author}{\bibfnamefont{S.~S.} \bibnamefont{Gubser}},
  \bibinfo{author}{\bibfnamefont{J.~M.} \bibnamefont{Maldacena}},
  \bibinfo{author}{\bibfnamefont{H.}~\bibnamefont{Ooguri}}, \bibnamefont{and}
  \bibinfo{author}{\bibfnamefont{Y.}~\bibnamefont{Oz}},
  \bibinfo{journal}{Phys.Rept.} \textbf{\bibinfo{volume}{323}},
  \bibinfo{pages}{183} (\bibinfo{year}{2000}), \eprint{hep-th/9905111}.

\bibitem[{\citenamefont{Witten}(1998{\natexlab{b}})}]{Witten:1998zw}
\bibinfo{author}{\bibfnamefont{E.}~\bibnamefont{Witten}},
  \bibinfo{journal}{Adv.Theor.Math.Phys.} \textbf{\bibinfo{volume}{2}},
  \bibinfo{pages}{505} (\bibinfo{year}{1998}{\natexlab{b}}),
  \eprint{hep-th/9803131}.

\bibitem[{\citenamefont{Caron-Huot et~al.}(2006)\citenamefont{Caron-Huot,
  Kovtun, Moore, Starinets, and Yaffe}}]{CaronHuot:2006te}
\bibinfo{author}{\bibfnamefont{S.}~\bibnamefont{Caron-Huot}},
  \bibinfo{author}{\bibfnamefont{P.}~\bibnamefont{Kovtun}},
  \bibinfo{author}{\bibfnamefont{G.~D.} \bibnamefont{Moore}},
  \bibinfo{author}{\bibfnamefont{A.}~\bibnamefont{Starinets}},
  \bibnamefont{and} \bibinfo{author}{\bibfnamefont{L.~G.} \bibnamefont{Yaffe}},
  \bibinfo{journal}{JHEP} \textbf{\bibinfo{volume}{0612}}, \bibinfo{pages}{015}
  (\bibinfo{year}{2006}), \eprint{hep-th/0607237}.

\bibitem[{\citenamefont{Mateos and Patino}(2007)}]{Mateos:2007yp}
\bibinfo{author}{\bibfnamefont{D.}~\bibnamefont{Mateos}} \bibnamefont{and}
  \bibinfo{author}{\bibfnamefont{L.}~\bibnamefont{Patino}},
  \bibinfo{journal}{JHEP} \textbf{\bibinfo{volume}{0711}}, \bibinfo{pages}{025}
  (\bibinfo{year}{2007}), \eprint{0709.2168}.

\bibitem[{\citenamefont{Sakai and Sugimoto}(2005)}]{Sakai:2004cn}
\bibinfo{author}{\bibfnamefont{T.}~\bibnamefont{Sakai}} \bibnamefont{and}
  \bibinfo{author}{\bibfnamefont{S.}~\bibnamefont{Sugimoto}},
  \bibinfo{journal}{Prog.Theor.Phys.} \textbf{\bibinfo{volume}{113}},
  \bibinfo{pages}{843} (\bibinfo{year}{2005}), \eprint{hep-th/0412141}.

\bibitem[{\citenamefont{Parnachev and Sahakyan}(2007)}]{Parnachev:2006ev}
\bibinfo{author}{\bibfnamefont{A.}~\bibnamefont{Parnachev}} \bibnamefont{and}
  \bibinfo{author}{\bibfnamefont{D.~A.} \bibnamefont{Sahakyan}},
  \bibinfo{journal}{Nucl.Phys.} \textbf{\bibinfo{volume}{B768}},
  \bibinfo{pages}{177} (\bibinfo{year}{2007}), \eprint{hep-th/0610247}.

\bibitem[{\citenamefont{Jo and Sin}(2011)}]{Jo:2010sg}
\bibinfo{author}{\bibfnamefont{K.}~\bibnamefont{Jo}} \bibnamefont{and}
  \bibinfo{author}{\bibfnamefont{S.-J.} \bibnamefont{Sin}},
  \bibinfo{journal}{Phys.Rev.} \textbf{\bibinfo{volume}{D83}},
  \bibinfo{pages}{026004} (\bibinfo{year}{2011}), \eprint{1005.0200}.

\bibitem[{\citenamefont{Bu}(2012)}]{PhysRevD.86.026003}
\bibinfo{author}{\bibfnamefont{Y.~Y.} \bibnamefont{Bu}},
  \bibinfo{journal}{Phys. Rev. D} \textbf{\bibinfo{volume}{86}},
  \bibinfo{pages}{026003} (\bibinfo{year}{2012}),
  \urlprefix\url{http://link.aps.org/doi/10.1103/PhysRevD.86.026003}.

\bibitem[{\citenamefont{Hassanain and Schvellinger}(2012)}]{Hassanain:2011ce}
\bibinfo{author}{\bibfnamefont{B.}~\bibnamefont{Hassanain}} \bibnamefont{and}
  \bibinfo{author}{\bibfnamefont{M.}~\bibnamefont{Schvellinger}},
  \bibinfo{journal}{Phys.Rev.} \textbf{\bibinfo{volume}{D85}},
  \bibinfo{pages}{086007} (\bibinfo{year}{2012}), \eprint{1110.0526}.

\bibitem[{\citenamefont{Baier et~al.}(2012{\natexlab{a}})\citenamefont{Baier,
  Stricker, Taanila, and Vuorinen}}]{Baier:2012tc}
\bibinfo{author}{\bibfnamefont{R.}~\bibnamefont{Baier}},
  \bibinfo{author}{\bibfnamefont{S.~A.} \bibnamefont{Stricker}},
  \bibinfo{author}{\bibfnamefont{O.}~\bibnamefont{Taanila}}, \bibnamefont{and}
  \bibinfo{author}{\bibfnamefont{A.}~\bibnamefont{Vuorinen}},
  \bibinfo{journal}{JHEP} \textbf{\bibinfo{volume}{1207}}, \bibinfo{pages}{094}
  (\bibinfo{year}{2012}{\natexlab{a}}), \eprint{1205.2998}.

\bibitem[{\citenamefont{Baier et~al.}(2012{\natexlab{b}})\citenamefont{Baier,
  Stricker, Taanila, and Vuorinen}}]{Baier:2012ax}
\bibinfo{author}{\bibfnamefont{R.}~\bibnamefont{Baier}},
  \bibinfo{author}{\bibfnamefont{S.~A.} \bibnamefont{Stricker}},
  \bibinfo{author}{\bibfnamefont{O.}~\bibnamefont{Taanila}}, \bibnamefont{and}
  \bibinfo{author}{\bibfnamefont{A.}~\bibnamefont{Vuorinen}}
  (\bibinfo{year}{2012}{\natexlab{b}}), \eprint{1207.1116}.

\bibitem[{\citenamefont{Steineder et~al.}(2012)\citenamefont{Steineder,
  Stricker, and Vuorinen}}]{Steineder:2012si}
\bibinfo{author}{\bibfnamefont{D.}~\bibnamefont{Steineder}},
  \bibinfo{author}{\bibfnamefont{S.~A.} \bibnamefont{Stricker}},
  \bibnamefont{and} \bibinfo{author}{\bibfnamefont{A.}~\bibnamefont{Vuorinen}}
  (\bibinfo{year}{2012}), \eprint{1209.0291}.

\bibitem[{\citenamefont{Steineder et~al.}(2013)\citenamefont{Steineder,
  Stricker, and Vuorinen}}]{Steineder:2013ana}
\bibinfo{author}{\bibfnamefont{D.}~\bibnamefont{Steineder}},
  \bibinfo{author}{\bibfnamefont{S.~A.} \bibnamefont{Stricker}},
  \bibnamefont{and} \bibinfo{author}{\bibfnamefont{A.}~\bibnamefont{Vuorinen}}
  (\bibinfo{year}{2013}), \eprint{1304.3404}.

\bibitem[{\citenamefont{van Hees et~al.}(2011)\citenamefont{van Hees, Gale, and
  Rapp}}]{vanHees:2011vb}
\bibinfo{author}{\bibfnamefont{H.}~\bibnamefont{van Hees}},
  \bibinfo{author}{\bibfnamefont{C.}~\bibnamefont{Gale}}, \bibnamefont{and}
  \bibinfo{author}{\bibfnamefont{R.}~\bibnamefont{Rapp}},
  \bibinfo{journal}{Phys.Rev.} \textbf{\bibinfo{volume}{C84}},
  \bibinfo{pages}{054906} (\bibinfo{year}{2011}), \eprint{1108.2131}.

\bibitem[{\citenamefont{Schenke and Strickland}(2007)}]{Schenke:2006yp}
\bibinfo{author}{\bibfnamefont{B.}~\bibnamefont{Schenke}} \bibnamefont{and}
  \bibinfo{author}{\bibfnamefont{M.}~\bibnamefont{Strickland}},
  \bibinfo{journal}{Phys.Rev.} \textbf{\bibinfo{volume}{D76}},
  \bibinfo{pages}{025023} (\bibinfo{year}{2007}), \eprint{hep-ph/0611332}.

\bibitem[{\citenamefont{Tuchin}(2011)}]{Tuchin:2010gx}
\bibinfo{author}{\bibfnamefont{K.}~\bibnamefont{Tuchin}},
  \bibinfo{journal}{Phys.Rev.} \textbf{\bibinfo{volume}{C83}},
  \bibinfo{pages}{017901} (\bibinfo{year}{2011}), \eprint{1008.1604}.

\bibitem[{\citenamefont{Tuchin}(2012)}]{Tuchin:2012mf}
\bibinfo{author}{\bibfnamefont{K.}~\bibnamefont{Tuchin}}
  (\bibinfo{year}{2012}), \eprint{1206.0485}.

\bibitem[{\citenamefont{Basar et~al.}(2012)\citenamefont{Basar, Kharzeev,
  Kharzeev, and Skokov}}]{Basar:2012bp}
\bibinfo{author}{\bibfnamefont{G.}~\bibnamefont{Basar}},
  \bibinfo{author}{\bibfnamefont{D.}~\bibnamefont{Kharzeev}},
  \bibinfo{author}{\bibfnamefont{D.}~\bibnamefont{Kharzeev}}, \bibnamefont{and}
  \bibinfo{author}{\bibfnamefont{V.}~\bibnamefont{Skokov}},
  \bibinfo{journal}{Phys.Rev.Lett.} \textbf{\bibinfo{volume}{109}},
  \bibinfo{pages}{202303} (\bibinfo{year}{2012}), \eprint{1206.1334}.

\bibitem[{\citenamefont{Fukushima and Mameda}(2012)}]{Fukushima:2012fg}
\bibinfo{author}{\bibfnamefont{K.}~\bibnamefont{Fukushima}} \bibnamefont{and}
  \bibinfo{author}{\bibfnamefont{K.}~\bibnamefont{Mameda}},
  \bibinfo{journal}{Phys.Rev.} \textbf{\bibinfo{volume}{D86}},
  \bibinfo{pages}{071501} (\bibinfo{year}{2012}), \eprint{1206.3128}.

\bibitem[{\citenamefont{Bzdak and Skokov}(2012)}]{Bzdak:2012fr}
\bibinfo{author}{\bibfnamefont{A.}~\bibnamefont{Bzdak}} \bibnamefont{and}
  \bibinfo{author}{\bibfnamefont{V.}~\bibnamefont{Skokov}}
  (\bibinfo{year}{2012}), \eprint{1208.5502}.

\bibitem[{\citenamefont{Rebhan and Steineder}(2011)}]{Rebhan:2011ke}
\bibinfo{author}{\bibfnamefont{A.}~\bibnamefont{Rebhan}} \bibnamefont{and}
  \bibinfo{author}{\bibfnamefont{D.}~\bibnamefont{Steineder}},
  \bibinfo{journal}{JHEP} \textbf{\bibinfo{volume}{1108}}, \bibinfo{pages}{153}
  (\bibinfo{year}{2011}), \eprint{1106.3539}.

\bibitem[{\citenamefont{Patino and Trancanelli}(2013)}]{Patino:2012py}
\bibinfo{author}{\bibfnamefont{L.}~\bibnamefont{Patino}} \bibnamefont{and}
  \bibinfo{author}{\bibfnamefont{D.}~\bibnamefont{Trancanelli}},
  \bibinfo{journal}{JHEP} \textbf{\bibinfo{volume}{1302}}, \bibinfo{pages}{154}
  (\bibinfo{year}{2013}), \eprint{1211.2199}.

\bibitem[{\citenamefont{Mateos and
  Trancanelli}(2011{\natexlab{a}})}]{Mateos:2011ix}
\bibinfo{author}{\bibfnamefont{D.}~\bibnamefont{Mateos}} \bibnamefont{and}
  \bibinfo{author}{\bibfnamefont{D.}~\bibnamefont{Trancanelli}},
  \bibinfo{journal}{Phys.Rev.Lett.} \textbf{\bibinfo{volume}{107}},
  \bibinfo{pages}{101601} (\bibinfo{year}{2011}{\natexlab{a}}),
  \eprint{1105.3472}.

\bibitem[{\citenamefont{Mateos and
  Trancanelli}(2011{\natexlab{b}})}]{Mateos:2011tv}
\bibinfo{author}{\bibfnamefont{D.}~\bibnamefont{Mateos}} \bibnamefont{and}
  \bibinfo{author}{\bibfnamefont{D.}~\bibnamefont{Trancanelli}},
  \bibinfo{journal}{JHEP} \textbf{\bibinfo{volume}{1107}}, \bibinfo{pages}{054}
  (\bibinfo{year}{2011}{\natexlab{b}}), \eprint{1106.1637}.

\bibitem[{\citenamefont{Bu}(2013)}]{PhysRevD.87.026005}
\bibinfo{author}{\bibfnamefont{Y.}~\bibnamefont{Bu}}, \bibinfo{journal}{Phys.
  Rev. D} \textbf{\bibinfo{volume}{87}}, \bibinfo{pages}{026005}
  (\bibinfo{year}{2013}),
  \urlprefix\url{http://link.aps.org/doi/10.1103/PhysRevD.87.026005}.

\bibitem[{\citenamefont{Yee}(2013)}]{Yee:2013qma}
\bibinfo{author}{\bibfnamefont{H.-U.} \bibnamefont{Yee}}
  (\bibinfo{year}{2013}), \eprint{1303.3571}.

\bibitem[{\citenamefont{Karch and Randall}(2001)}]{Karch:2000gx}
\bibinfo{author}{\bibfnamefont{A.}~\bibnamefont{Karch}} \bibnamefont{and}
  \bibinfo{author}{\bibfnamefont{L.}~\bibnamefont{Randall}},
  \bibinfo{journal}{JHEP} \textbf{\bibinfo{volume}{0106}}, \bibinfo{pages}{063}
  (\bibinfo{year}{2001}), \eprint{hep-th/0105132}.

\bibitem[{\citenamefont{Karch and Katz}(2002)}]{Karch:2002sh}
\bibinfo{author}{\bibfnamefont{A.}~\bibnamefont{Karch}} \bibnamefont{and}
  \bibinfo{author}{\bibfnamefont{E.}~\bibnamefont{Katz}},
  \bibinfo{journal}{JHEP} \textbf{\bibinfo{volume}{0206}}, \bibinfo{pages}{043}
  (\bibinfo{year}{2002}), \eprint{hep-th/0205236}.

\bibitem[{\citenamefont{Son and Starinets}(2002)}]{Son:2002sd}
\bibinfo{author}{\bibfnamefont{D.~T.} \bibnamefont{Son}} \bibnamefont{and}
  \bibinfo{author}{\bibfnamefont{A.~O.} \bibnamefont{Starinets}},
  \bibinfo{journal}{JHEP} \textbf{\bibinfo{volume}{0209}}, \bibinfo{pages}{042}
  (\bibinfo{year}{2002}), \eprint{hep-th/0205051}.

\bibitem[{\citenamefont{Kovtun and Starinets}(2005)}]{Kovtun:2005ev}
\bibinfo{author}{\bibfnamefont{P.~K.} \bibnamefont{Kovtun}} \bibnamefont{and}
  \bibinfo{author}{\bibfnamefont{A.~O.} \bibnamefont{Starinets}},
  \bibinfo{journal}{Phys.Rev.} \textbf{\bibinfo{volume}{D72}},
  \bibinfo{pages}{086009} (\bibinfo{year}{2005}), \eprint{hep-th/0506184}.

\bibitem[{\citenamefont{Mateos et~al.}(2006)\citenamefont{Mateos, Myers, and
  Thomson}}]{Mateos:2006nu}
\bibinfo{author}{\bibfnamefont{D.}~\bibnamefont{Mateos}},
  \bibinfo{author}{\bibfnamefont{R.~C.} \bibnamefont{Myers}}, \bibnamefont{and}
  \bibinfo{author}{\bibfnamefont{R.~M.} \bibnamefont{Thomson}},
  \bibinfo{journal}{Phys.Rev.Lett.} \textbf{\bibinfo{volume}{97}},
  \bibinfo{pages}{091601} (\bibinfo{year}{2006}), \eprint{hep-th/0605046}.

\bibitem[{\citenamefont{Hoyos-Badajoz et~al.}(2007)\citenamefont{Hoyos-Badajoz,
  Landsteiner, and Montero}}]{Hoyos:2006gb}
\bibinfo{author}{\bibfnamefont{C.}~\bibnamefont{Hoyos-Badajoz}},
  \bibinfo{author}{\bibfnamefont{K.}~\bibnamefont{Landsteiner}},
  \bibnamefont{and} \bibinfo{author}{\bibfnamefont{S.}~\bibnamefont{Montero}},
  \bibinfo{journal}{JHEP} \textbf{\bibinfo{volume}{0704}}, \bibinfo{pages}{031}
  (\bibinfo{year}{2007}), \eprint{hep-th/0612169}.

\bibitem[{\citenamefont{Mateos et~al.}(2007)\citenamefont{Mateos, Myers, and
  Thomson}}]{Mateos:2007vn}
\bibinfo{author}{\bibfnamefont{D.}~\bibnamefont{Mateos}},
  \bibinfo{author}{\bibfnamefont{R.~C.} \bibnamefont{Myers}}, \bibnamefont{and}
  \bibinfo{author}{\bibfnamefont{R.~M.} \bibnamefont{Thomson}},
  \bibinfo{journal}{JHEP} \textbf{\bibinfo{volume}{0705}}, \bibinfo{pages}{067}
  (\bibinfo{year}{2007}), \eprint{hep-th/0701132}.

\bibitem[{\citenamefont{Wu et~al.}(in progress)\citenamefont{Wu, Yang, and
  Yang}}]{wuyang}
\bibinfo{author}{\bibfnamefont{S.-Y.} \bibnamefont{Wu}},
  \bibinfo{author}{\bibfnamefont{D.-L.} \bibnamefont{Yang}}, \bibnamefont{and}
  \bibinfo{author}{\bibfnamefont{Y.}~\bibnamefont{Yang}} (\bibinfo{year}{in
  progress}).

\bibitem[{\citenamefont{Filev et~al.}(2007)\citenamefont{Filev, Johnson,
  Rashkov, and Viswanathan}}]{Filev:2007gb}
\bibinfo{author}{\bibfnamefont{V.~G.} \bibnamefont{Filev}},
  \bibinfo{author}{\bibfnamefont{C.~V.} \bibnamefont{Johnson}},
  \bibinfo{author}{\bibfnamefont{R.}~\bibnamefont{Rashkov}}, \bibnamefont{and}
  \bibinfo{author}{\bibfnamefont{K.}~\bibnamefont{Viswanathan}},
  \bibinfo{journal}{JHEP} \textbf{\bibinfo{volume}{0710}}, \bibinfo{pages}{019}
  (\bibinfo{year}{2007}), \eprint{hep-th/0701001}.

\bibitem[{\citenamefont{Albash et~al.}(2008{\natexlab{a}})\citenamefont{Albash,
  Filev, Johnson, and Kundu}}]{Albash:2007bk}
\bibinfo{author}{\bibfnamefont{T.}~\bibnamefont{Albash}},
  \bibinfo{author}{\bibfnamefont{V.~G.} \bibnamefont{Filev}},
  \bibinfo{author}{\bibfnamefont{C.~V.} \bibnamefont{Johnson}},
  \bibnamefont{and} \bibinfo{author}{\bibfnamefont{A.}~\bibnamefont{Kundu}},
  \bibinfo{journal}{JHEP} \textbf{\bibinfo{volume}{0807}}, \bibinfo{pages}{080}
  (\bibinfo{year}{2008}{\natexlab{a}}), \eprint{0709.1547}.

\bibitem[{\citenamefont{Albash et~al.}(2008{\natexlab{b}})\citenamefont{Albash,
  Filev, Johnson, and Kundu}}]{Albash:2007bq}
\bibinfo{author}{\bibfnamefont{T.}~\bibnamefont{Albash}},
  \bibinfo{author}{\bibfnamefont{V.~G.} \bibnamefont{Filev}},
  \bibinfo{author}{\bibfnamefont{C.~V.} \bibnamefont{Johnson}},
  \bibnamefont{and} \bibinfo{author}{\bibfnamefont{A.}~\bibnamefont{Kundu}},
  \bibinfo{journal}{JHEP} \textbf{\bibinfo{volume}{0808}}, \bibinfo{pages}{092}
  (\bibinfo{year}{2008}{\natexlab{b}}), \eprint{0709.1554}.

\bibitem[{\citenamefont{Tuchin}(2013{\natexlab{a}})}]{Tuchin:2013ie}
\bibinfo{author}{\bibfnamefont{K.}~\bibnamefont{Tuchin}}
  (\bibinfo{year}{2013}{\natexlab{a}}), \eprint{1301.0099}.

\bibitem[{\citenamefont{Tuchin}(2013{\natexlab{b}})}]{Tuchin:2013apa}
\bibinfo{author}{\bibfnamefont{K.}~\bibnamefont{Tuchin}}
  (\bibinfo{year}{2013}{\natexlab{b}}), \eprint{1305.5806}.

\bibitem[{\citenamefont{McLerran and Skokov}(2013)}]{McLerran:2013hla}
\bibinfo{author}{\bibfnamefont{L.}~\bibnamefont{McLerran}} \bibnamefont{and}
  \bibinfo{author}{\bibfnamefont{V.}~\bibnamefont{Skokov}}
  (\bibinfo{year}{2013}), \eprint{1305.0774}.

\bibitem[{\citenamefont{Casalderrey-Solana and
  Mateos}(2009)}]{CasalderreySolana:2008ne}
\bibinfo{author}{\bibfnamefont{J.}~\bibnamefont{Casalderrey-Solana}}
  \bibnamefont{and} \bibinfo{author}{\bibfnamefont{D.}~\bibnamefont{Mateos}},
  \bibinfo{journal}{Phys.Rev.Lett.} \textbf{\bibinfo{volume}{102}},
  \bibinfo{pages}{192302} (\bibinfo{year}{2009}), \eprint{0806.4172}.

\bibitem[{\citenamefont{Giataganas}(2012)}]{Giataganas:2012zy}
\bibinfo{author}{\bibfnamefont{D.}~\bibnamefont{Giataganas}},
  \bibinfo{journal}{JHEP} \textbf{\bibinfo{volume}{1207}}, \bibinfo{pages}{031}
  (\bibinfo{year}{2012}), \eprint{1202.4436}.

\bibitem[{\citenamefont{Chernicoff et~al.}(2013)\citenamefont{Chernicoff,
  Fernandez, Mateos, and Trancanelli}}]{Chernicoff:2012bu}
\bibinfo{author}{\bibfnamefont{M.}~\bibnamefont{Chernicoff}},
  \bibinfo{author}{\bibfnamefont{D.}~\bibnamefont{Fernandez}},
  \bibinfo{author}{\bibfnamefont{D.}~\bibnamefont{Mateos}}, \bibnamefont{and}
  \bibinfo{author}{\bibfnamefont{D.}~\bibnamefont{Trancanelli}},
  \bibinfo{journal}{JHEP} \textbf{\bibinfo{volume}{1301}}, \bibinfo{pages}{170}
  (\bibinfo{year}{2013}), \eprint{1208.2672}.

\bibitem[{\citenamefont{Chakraborty and Haque}(2012)}]{Chakraborty:2012dt}
\bibinfo{author}{\bibfnamefont{S.}~\bibnamefont{Chakraborty}} \bibnamefont{and}
  \bibinfo{author}{\bibfnamefont{N.}~\bibnamefont{Haque}}
  (\bibinfo{year}{2012}), \eprint{1212.2769}.

\bibitem[{\citenamefont{Dumitru et~al.}(2008)\citenamefont{Dumitru, Guo, and
  Strickland}}]{Dumitru:2007hy}
\bibinfo{author}{\bibfnamefont{A.}~\bibnamefont{Dumitru}},
  \bibinfo{author}{\bibfnamefont{Y.}~\bibnamefont{Guo}}, \bibnamefont{and}
  \bibinfo{author}{\bibfnamefont{M.}~\bibnamefont{Strickland}},
  \bibinfo{journal}{Phys.Lett.} \textbf{\bibinfo{volume}{B662}},
  \bibinfo{pages}{37} (\bibinfo{year}{2008}), \eprint{0711.4722}.

\bibitem[{\citenamefont{Dumitru et~al.}(2009)\citenamefont{Dumitru, Guo, and
  Strickland}}]{Dumitru:2009fy}
\bibinfo{author}{\bibfnamefont{A.}~\bibnamefont{Dumitru}},
  \bibinfo{author}{\bibfnamefont{Y.}~\bibnamefont{Guo}}, \bibnamefont{and}
  \bibinfo{author}{\bibfnamefont{M.}~\bibnamefont{Strickland}},
  \bibinfo{journal}{Phys.Rev.} \textbf{\bibinfo{volume}{D79}},
  \bibinfo{pages}{114003} (\bibinfo{year}{2009}), \eprint{0903.4703}.

\bibitem[{\citenamefont{Strickland and Bazow}(2012)}]{Strickland:2011aa}
\bibinfo{author}{\bibfnamefont{M.}~\bibnamefont{Strickland}} \bibnamefont{and}
  \bibinfo{author}{\bibfnamefont{D.}~\bibnamefont{Bazow}},
  \bibinfo{journal}{Nucl.Phys.} \textbf{\bibinfo{volume}{A879}},
  \bibinfo{pages}{25} (\bibinfo{year}{2012}), \eprint{1112.2761}.

\bibitem[{\citenamefont{Strickland}(2011)}]{Strickland:2011mw}
\bibinfo{author}{\bibfnamefont{M.}~\bibnamefont{Strickland}},
  \bibinfo{journal}{Phys.Rev.Lett.} \textbf{\bibinfo{volume}{107}},
  \bibinfo{pages}{132301} (\bibinfo{year}{2011}), \eprint{1106.2571}.

\bibitem[{\citenamefont{Mocsy et~al.}(2013)\citenamefont{Mocsy, Petreczky, and
  Strickland}}]{Mocsy:2013syh}
\bibinfo{author}{\bibfnamefont{A.}~\bibnamefont{Mocsy}},
  \bibinfo{author}{\bibfnamefont{P.}~\bibnamefont{Petreczky}},
  \bibnamefont{and}
  \bibinfo{author}{\bibfnamefont{M.}~\bibnamefont{Strickland}},
  \bibinfo{journal}{Int.J.Mod.Phys.} \textbf{\bibinfo{volume}{A28}},
  \bibinfo{pages}{1340012} (\bibinfo{year}{2013}), \eprint{1302.2180}.

\bibitem[{\citenamefont{Rebhan and Steineder}(2012)}]{Rebhan:2011vd}
\bibinfo{author}{\bibfnamefont{A.}~\bibnamefont{Rebhan}} \bibnamefont{and}
  \bibinfo{author}{\bibfnamefont{D.}~\bibnamefont{Steineder}},
  \bibinfo{journal}{Phys.Rev.Lett.} \textbf{\bibinfo{volume}{108}},
  \bibinfo{pages}{021601} (\bibinfo{year}{2012}), \eprint{1110.6825}.

\bibitem[{\citenamefont{Muller et~al.}(to appear)\citenamefont{Muller, Wu, and
  Yang}}]{wuyangv2}
\bibinfo{author}{\bibfnamefont{B.}~\bibnamefont{Muller}},
  \bibinfo{author}{\bibfnamefont{S.-Y.} \bibnamefont{Wu}}, \bibnamefont{and}
  \bibinfo{author}{\bibfnamefont{D.-L.} \bibnamefont{Yang}} (\bibinfo{year}{to
  appear}).

\end{thebibliography}

\end{document}